\documentclass{IEEEtran}

\usepackage[noadjust]{cite}

\usepackage[T1]{fontenc}
\usepackage{newtxtext}
\usepackage[cmintegrals]{newtxmath}
\usepackage{bm}
\usepackage{amsmath,amsfonts}

\usepackage{graphicx}
\usepackage[caption=false,font=footnotesize]{subfig}
\usepackage{array}
\usepackage{float}
\usepackage{wrapfig} 

\usepackage{xcolor}
\usepackage{textcomp}
\usepackage{url}
\usepackage{balance}
\usepackage{algorithm}
\usepackage{algpseudocode}

\usepackage[hypertexnames=true, bookmarks=true, bookmarksnumbered=true,
pdfpagemode=UseOutlines, plainpages=false, pdfpagelabels=true,
colorlinks=true, linkcolor=black, citecolor=black, urlcolor=black]{hyperref}

\begin{document}

\title{Unlicensed Band Allocation for Heterogeneous Networks}

\author{Po-Heng Chou,~\IEEEmembership{Member,~IEEE} \vspace{-0.4in}

\thanks{P.-H. Chou is with the Graduate Institute of Communication Engineering (GICE), National Taiwan University (NTU), Taipei 10617, Taiwan (e-mail:d00942015@ntu.edu.tw).}
\thanks{This work was supported in part by the Ministry of Science and
Technology (MOST) of Taiwan under Grant MOST 103-2218-E-002-032, 104-3115-E-002-004, 105-2218-E-002-013, and 106-2218-E-002-029.}
}

{}

\maketitle
\begin{abstract}
Based on the License-Assisted Access (LAA) small cell architecture, the LAA coexisting with Wi-Fi heterogeneous networks provides LTE mobile users with high bandwidth efficiency as the unlicensed channels are shared among LAA and Wi-Fi. However, LAA and Wi-Fi interfere with each other when both systems use the same unlicensed channel in heterogeneous networks. In such a network, unlicensed band allocation for LAA and Wi-Fi is an important issue that may affect the quality of service (QoS) of both systems significantly. In this paper, we propose an analytical model and conduct simulation experiments to study four allocations for the unlicensed band: unlicensed full allocation (UFA), unlicensed time-division allocation (UTA), and UFA/UTA with buffering mechanism (UFAB and UTAB) for the LAA data packets. We evaluate the performance of these unlicensed band allocation schemes in terms of the acceptance rate of both LAA and Wi-Fi packet data in the LAA buffer queue. Our study provides guidelines for designing the channel occupation phase and the buffer size of the LAA small cell.
\end{abstract}


\IEEEpeerreviewmaketitle

\section{Introduction}
\label{sect:introduction}

The diversity of wireless Internet applications has driven demands for more bandwidth in the 5G networks~\cite{5G2015}. Utilization of unlicensed bands has been
considered as one of the promising technologies for realizing more bandwidth.
For example, License-Assisted Access
(LAA)~\cite{3GPP2015} and LTE-Unlicensed
(LTE-U)~\cite{LTE-U2015,LTE-U_Qualcomm2014} have been proposed by the 3rd
Generation Partnership Project (3GPP) and LTE-U Forum to allow LTE
to utilize the unlicensed band that is also being used by the incumbent
systems (e.g., Wi-Fi), respectively.
The main difference between LAA and LTE-U lies in their inter-system interference avoidance mechanisms.
The LAA adopts the Listen Before Talk (LBT) mechanism, where the LTE base station (eNB)
periodically senses the target unlicensed band to determine whether the unlicensed band is idle or not, and begins the data
transmission if the target unlicensed band is idle.
The LTE-U adopts the Carrier Sense Adaptive Transmission (CSAT) mechanism.
In CSAT, a time-division-multiplexing (TDM) cycle
is defined, which consists of an LTE-U\_On period and an LTE-U\_Off period.
In the LTE-U\_On period, the LTE eNB transmits data through the unlicensed band.
On the other hand, in the LTE-U\_Off period, the LTE eNB transmits data through the licensed band.
As pointed out in~\cite{Leith2015,Andra2016,Quek2017,Bolin2017}, the CSAT has the LTE system compete in the unlicensed bands more aggressively with the incumbent system,
which results in unfair unlicensed band allocation to the LTE and the incumbent system.
Furthermore, the overhead (in terms of the channel time spent resolving LTE and Wi-Fi collisions when accessing unlicensed bands) of
CSAT is higher than that of LBT.
In this paper, we consider the LAA mechanism to study the performance of the unlicensed band allocation in heterogeneous networks.

To more efficiently utilize the unlicensed band, previous studies~\cite{SmallCell2012,Liu2011,Liu2015} suggested the adoption of the small cell technology\footnote[2]{The small cell provides wireless
transmission with a high data rate, offloads data traffic from a macro
cell, and extends the service area of a macro cell, i.e., outdoor BSs
run LTE or Wi-Fi protocols on the unlicensed band.} to run the LTE protocol or the Wi-Fi protocol on the unlicensed bands.
If Wi-Fi is adopted in the small cells, the handover from the LTE macro cell to the small cell becomes an inter-system handover, which incurs long delays and is more prone to failure.
Details of the handover procedure between the LTE macro cell and the Wi-Fi small cell can be found in~\cite{E-UTRAN2015}.
To resolve the issue, it is preferred that the small cell runs the LTE protocol, and the handover between the
LTE macro cell and the LTE small cell is an intra-system handover.

To mitigate the intra-system interference imposed by the LAA small
cell on the LTE macro cell, the previous works focused on
unlicensed band allocation for LAA small cells. Several technical
reports~\cite{5G2015,Zhang2015,Liu2015,LTE-U_Qualcomm2014,LTE-U_Qualcomm2013,LTE-U_Cisco2013}
suggest that LAA should run on the Wi-Fi unlicensed band (e.g.,
$5.15-5.25$ GHz Indoor/outdoor in the US~\cite{5G2015}).
However, the heavy Wi-Fi traffic loads in the unlicensed bands may result in the LAA small cells suffering significant interference~\cite{5G2015,Zhang2015}.
As pointed out in~\cite{LTE-U_Qualcomm2014}, the LAA small cell is a dual-band small cell that simultaneously uses both the licensed and unlicensed bands through the LTE carrier aggregation (CA)~\cite{Zhang2014}.
The transmission on the unlicensed band is usually unstable, which may not offer transmission with guaranteed QoS.
The design of the LAA channel access mechanism and unlicensed band allocation methods to ensure
well-balanced coexistence of LAA small cells and Wi-Fi APs is one of the important issues, providing
both LAA and Wi-Fi fair opportunities to access the unlicensed band.

\begin{figure}[t]
\centering
\includegraphics[width=.4\textwidth]{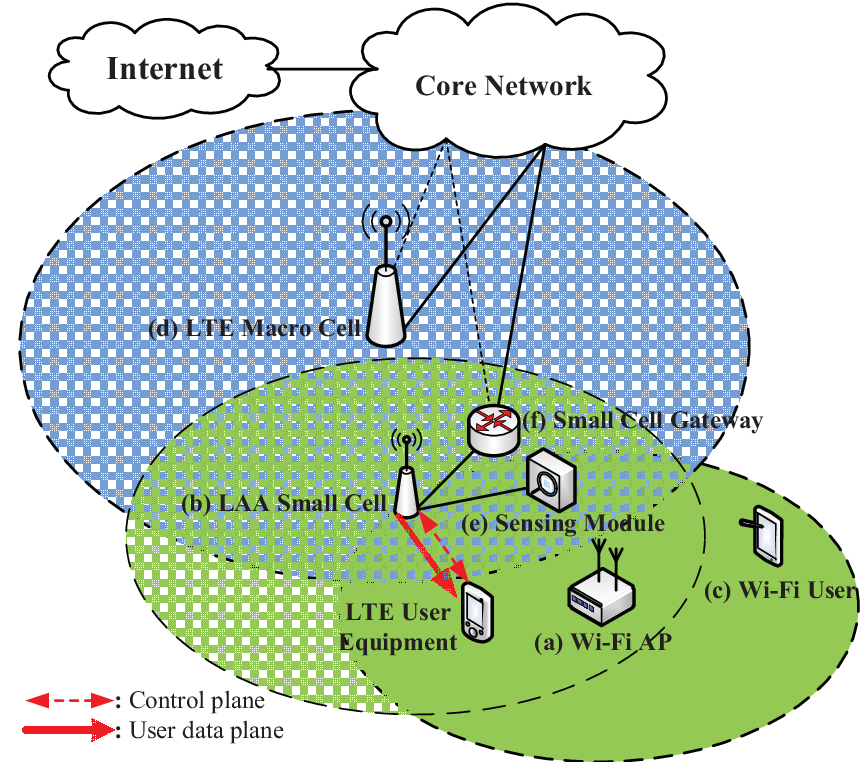}
\caption{The LAA small cell coexisting with Wi-Fi AP heterogeneous network architecture with unlicensed band.}
\label{Architecture}
\end{figure}

In this paper, we consider the coexistence scenario for LAA small cells and Wi-Fi APs.
Fig.~\ref{Architecture} shows a general architecture~\cite{5G2015,Zhang2015,Liu2015,Khawer2016} of the heterogeneous network consisting of an LTE macro cell, an LTE small cell (running LAA), and a Wi-Fi AP (i.e., the incumbent system).
The Wi-Fi AP [Fig.~\ref{Architecture} (a)] and the LAA small cell [Fig.~\ref{Architecture} (b)] use the unlicensed band.
The LAA small cell overlaps with an LTE macro cell [Fig.~\ref{Architecture} (d)] and the Wi-Fi AP.
A sensing module [Fig.~\ref{Architecture} (e)] is attached on the LAA small cell,
which is responsible for measuring the interference caused by the communication between the Wi-Fi APs and incumbent
users on the unlicensed bands and for controlling the LAA small cell to use the unlicensed band or not.
An example of the protocols between the sensing module and the LAA small cell can be found in~\cite{Sleep2011}.
The LAA small cell connects the LTE core network through the small cell gateway [SC-GW; Fig.~\ref{Architecture} (f)].

Two operation modes are defined for LAA~\cite{Bolin2017,LTE-U2015,Zhang2015,LTE-U_Qualcomm2014}: supplemental downlink (SDL) and time-division duplex (TDD).
Because the downlink traffic is heavier than the uplink traffic in most Internet applications, in this study, we focus on the SDL mode.
In the SDL mode, the unlicensed band is used to carry downlink traffic, while the uplink transmissions and control signals are transmitted on the licensed band.
\begin{figure}[t]
\centering
\includegraphics[width=.4\textwidth]{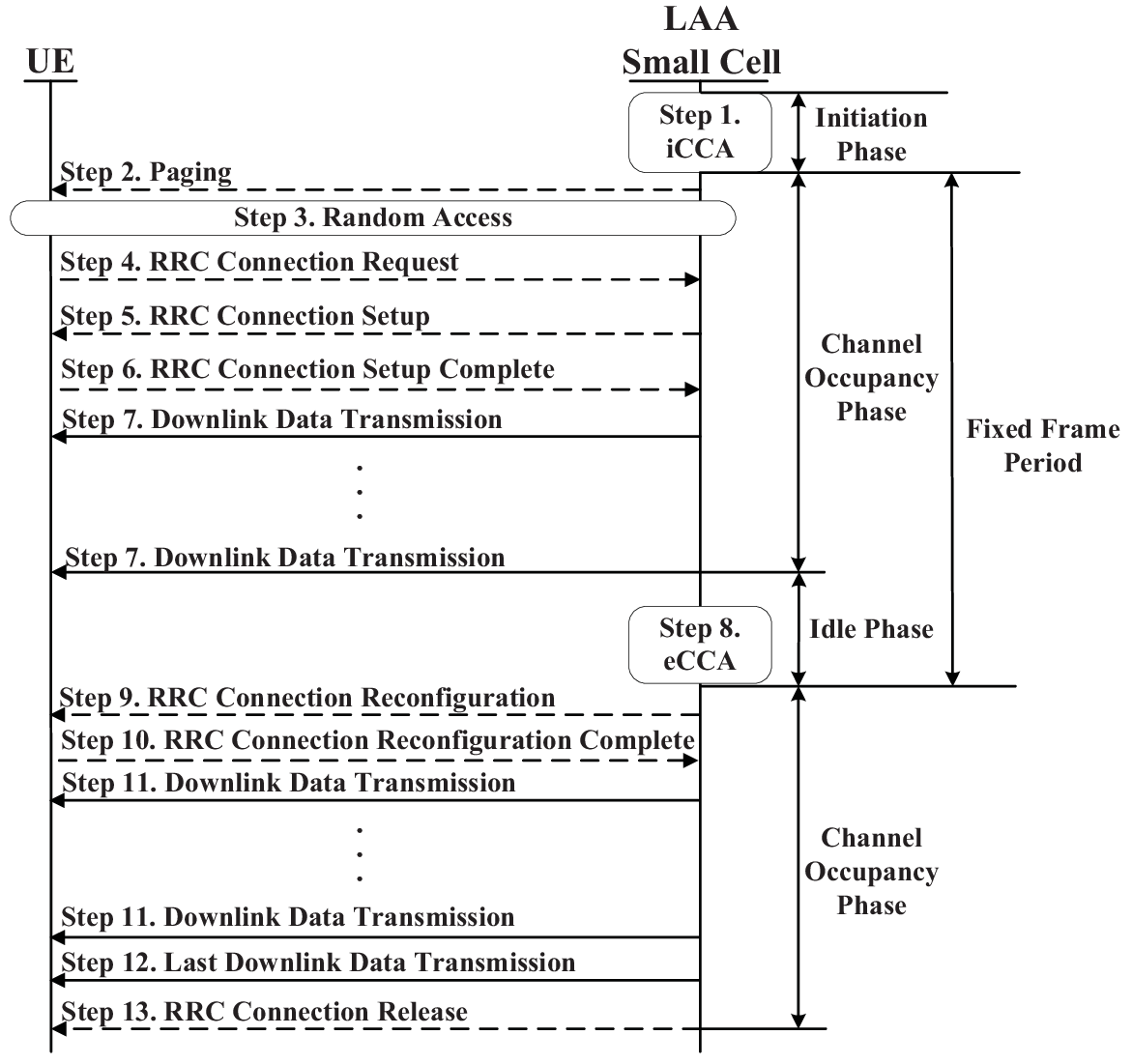}
\caption{LAA downlink packet transfer.}
\label{Flow}
\end{figure}
Fig.~\ref{Flow} illustrates the message flow for the LAA small cell in the SDL mode to have downlink transmission to a UE on the unlicensed bands~\cite{3GPP2015},
where the dashed lines represent the signaling message exchanges, and the solid lines represent the user data transmission.
There are three phases: the initiation phase, the channel occupancy phase, and the idle phase.
The channel occupancy phase and the idle phase consist of fixed numbers of frame periods.
Note that these numbers could be pre-configured by the manufacturers.
The channel occupancy phase is followed by the idle phase.
The time period of the idle phase is at least $5\%$ of the channel occupancy phase.
The details of the message flow are given below:
\begin{itemize}
\item \textbf{Step 1.} When an LAA small cell has a packet to transmit to the UE, it initiates the downlink packet transfer (i.e., in the initiation phase) by performing the LBT procedure, also known as the initiation clear channel assessment (iCCA)~\cite{3GPP2015}.
In the LBT procedure, the sensing module first senses the unlicensed band (i.e., unlicensed channel).
If the unlicensed channel is clear (i.e., idle channel), the LAA small cell starts transmission by executing Step 2.
Otherwise (i.e., there is interference on the unlicensed channel), the packet is queued at the LAA small cell.
The execution of the LBT procedure takes at least $20$ $\mu$s.

\item \textbf{Step 2.} The LAA small cell pages the UE through the control channel (i.e., on the licensed band),
where the resource allocation information is carried in the paging message.
When the UE receives the paging message, it determines whether its identity (ID) is carried in the paging message. If so, the UE executes Step 3.

\item \textbf{Step 3-6.} The UE triggers the random access procedure~\cite{Random_Access_Procedure}, and exchanges the radio resource control (RRC) connection request, Setup, and Complete messages to establish an RRC connection.
    Note that Steps 2-6 are the channel occupancy phase.

\item \textbf{Step 7.} The UE starts to download data blocks from the LAA small cell via the unlicensed band according to the agreed channel assignment (i.e., in the channel occupancy phase, at least $1$ $m$ sec, at most $10$ $m$ sec). The transmission lasts for at least $1$ ms and at most $10$ ms.
    Then the transmission moves to the idle phase.

\item \textbf{Step 8.} At the end of the idle phase, the LAA small cell performs the extended clear channel assessment (eCCA) for the next channel occupancy phase.

\item \textbf{Step 9-11.} Based on the result of the eCCA,
If the unlicensed channel is not clear, the LAA small cell may reallocate other idle channels for the UE according to the last eCCA results via the RRC connection reconfiguration procedure~\cite{RRC2015}, and the LAA small cell starts transmission on another idle unlicensed channel.
Otherwise (i.e., the unlicensed channel is still idle), Steps 9-10 are skipped, and the LAA small cell starts transmission by the same unlicensed channel (i.e., Step 11).

\item \textbf{Step 12-13.} When the LAA completes the downlink transmission to the UE, the information is carried in the last data block, and the LAA small cell then terminates the downlink transmission by returning the RRC connection release.
\end{itemize}

Note that at Step 9 (channel reassignment), there is no control for the LAA small cell to release the unlicensed channel, which may result in significant performance degradation for the Wi-Fi users.
For example, when LTE transmits the video streams, it will occupy the unlicensed channel until the end of playing the video; the Wi-Fi does not have an opportunity for transmission for a long time, and the Wi-Fi QoS can be significantly degraded.
To prevent such performance degradation, a channel allocation mechanism needs to be designed at Step 9.
In this paper, we use four unlicensed channel allocations, including
unlicensed full allocation (UFA) and unlicensed time-division allocation (UTA), unlicensed full allocation with buffering mechanism (UFAB), and unlicensed time-division allocation with buffering mechanism (UTAB) at the LAA small cell.
We propose an analytic model and simulation experiments to evaluate the performance of these allocations.
To the best of our knowledge, our analytical model is the first one that puts the traffic model, the finite buffer size, and the LBT-based (ON/OFF) behavior into consideration to study the performance of the LAA coexisting with Wi-Fi, i.e., the effects of the MAC layer interaction between the LAA and Wi-Fi.
We will survey the related works about modeling for LAA and Wi-Fi coexistence in the following section.

\section{Related Works}
\label{sect:related works}
This section surveys the previous works that studied the coexistence of LAA and Wi-Fi.
The works~\cite{Bolin2017,Andra2016} have surveyed and provided case studies for the LAA and Wi-Fi coexistence systems.
In \cite{Bolin2017,Andra2016}, they show that efficient unlicensed channel allocation for LAA and Wi-Fi coexistence is a critical issue that significantly affects performance and QoS.
Therefore, most of the previous studies focused on the fairness between the LAA and Wi-Fi.
The previous works~\cite{Liu2015,Cano2015,Chen2016,Qixun2016,Quek2017} that resolved the radio resource allocation issue in LAA coexisting with Wi-Fi are discussed below:
In~\cite{Liu2015}, the authors proposed a radio resource allocation scheme to balance the traffic over licensed bands and unlicensed bands.
The authors attempted to maintain the throughput of the Wi-Fi network by using a utility-based optimization framework.
In~\cite{Cano2015}, a novel proportional fair allocation scheme was proposed to ensure fair coexistence between LAA and Wi-Fi networks.
However, in~\cite{Liu2015,Cano2015}, only a single LAA UE is considered.
They did not touch on radio resource allocation for the scenario of multiple UEs, where the harmonious coexistence among all the devices should be addressed.
In~\cite{Chen2016}, a hybrid method combining traffic offloading and radio resource allocation methods was proposed to deliver cellular data traffic over an unlicensed band.
In~\cite{Qixun2016}, the authors proposed a radio resource scheduling approach.
The authors applied the linear programming algorithm, RSA-LP, to maximize the throughput of LAA and Wi-Fi networks.

However, the occupation of the LAA network will change the behavior of the Wi-Fi network, which was not considered in~\cite{Chen2016,Qixun2016}.
Therefore, in~\cite{Quek2017}, a cross-layer proportional fairness (PF)-based framework
to jointly optimize the protocol parameters of the MAC layer and the physical layer of an LAA network was proposed.
However, the duty cycle-based schemes reported so far do not allow for offering performance guarantees to the LTE connections, as all of the arriving packets are admitted, and the LTE traffic variation is made.
Furthermore, LBT mechanism optimization makes LTE complicated to employ one of the most attractive features of LAA, that is, provisioning of QoS guarantees.
When the Wi-Fi and LAA compete for the shared resources, no strict delay constraints can be met due to the intrinsic properties of the CSMA protocol family.
And the LBT mechanism as specified by 3GPP is fundamentally different from the CSMA/CA protocol in IEEE 802.11 systems by implying that it may not be trivial to identify the parameters that ensure fairness of resource allocation~\cite{Li2015}.
Since LAA BS can be granted full control of the shared spectrum, the competition between LTE and Wi-Fi may not be needed to achieve fairness.
Hence, instead of competing for resources with Wi-Fi, LAA BS may manage the resource allocation between Wi-Fi and LTE, thus dividing them in a fair manner while keeping track of the number of concurrent LTE packets and the state of the unlicensed band in proximity of the LAA BS.
The previous works~\cite{Qixun2016,Lee2018,Mehrnoush2018,Tang2018} considered the interactions between the LAA and the Wi-Fi networks by utilizing two Markov chains separately.
However, the interaction between the Markov model for LAA and that for Wi-Fi was modeled by only the collision probability,
which may not properly reflect the interaction in the real systems.
To address this issue, in this paper, we model the interaction between LAA and Wi-Fi by using a global system state.
Furthermore, the MAC layer performance studies in the previous works did not consider the traffic model in the application layer,
so the studies may not reflect the situation in the real system.

In this paper, we consider both the application traffic model (based on the
3GPP FTP traffic model~\cite{3GPP36.814}) in the analytical model and the simulation model.
This paper proposes an analytical model to study the coexisting issue for LAA small cells operating on the unlicensed band with Wi-Fi heterogeneous networks.
The model quantifies the length of the channel occupation phase of an LAA small cell, which is referred to as the ON/OFF service rate.
It is clear that the shorter the length of the channel occupation phase, the less it affects the Wi-Fi system.
However, reducing the length of the channel occupation phase may, at the same time, degrade the LAA system performance in terms of dropping probability.
Therefore, we also quantify the LAA packet dropping probability to examine the penalty caused by exercising the unlicensed band allocation for LAA/Wi-Fi heterogeneous networks.
Due to the complicated ON/OFF behaviors of an LAA small cell, the proposed analytical model may not well capture the LAA small cell ON/OFF behavior under some conditions.
To release these constraints of the analytical model, we conduct simulation experiments as well.
We note that the performance of the unlicensed band allocation primarily depends on the length of the sensing phase/channel occupation phase and the Wi-Fi arrival rates.
Based on the LBT-based behavior and buffering mechanism, we proposed four schemes for unlicensed band allocation in this paper, including the UFA, UFAB, UTA, and UTAB. And the UTAB has been mentioned in~\cite{Chou2019}.
Furthermore, our study indicates that with proper parameter settings, the buffer mechanism can reduce the effect of the Wi-Fi without significantly increasing the LAA packet dropping probability.
The analytical and simulation results of this work can serve as guidelines for the implementation of the LAA small cell.

The remaining parts of this paper are organized as follows.
Section~\ref{sect:allocation} introduces the UFA and UTA allocations and the corresponding buffering mechanism.
The proposed analytic model for resource allocations in the previous section is described in Section~\ref{sect:analytical}.
In Section~\ref{sect:performance}, we study the performances of UFA and UTA allocations and the corresponding buffering mechanism.
Finally, we conclude this work in Section~\ref{sect:conclusion}.

\section{Radio Resource Allocations}
\label{sect:allocation}

\begin{figure}[t]
\centering
\includegraphics[width=.30\textwidth]{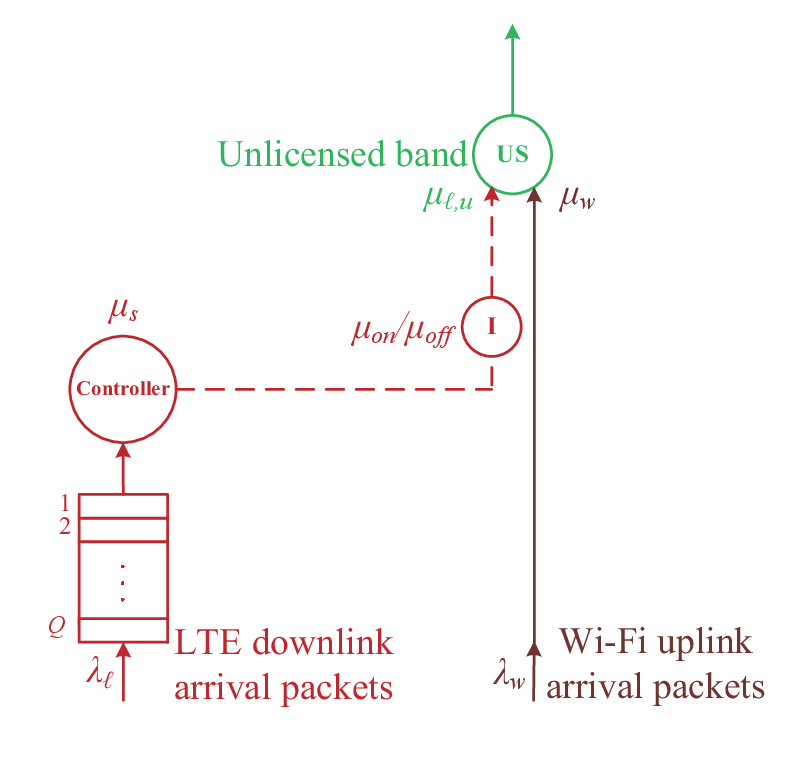}
\caption{The queueing model of LAA coexisting with Wi-Fi heterogeneous networks.}
\label{LAA/Wifi_Queuing}
\end{figure}

This section uses the two unlicensed channel allocations, UFA and UTA, and the buffering mechanism (which can be applied to UFA and UTA) for
An LAA small cell to allocate unlicensed channels for downloading packet transmission.
We show the UFA and UTA operation based on the queuing model for the LAA and Wi-Fi coexisting system shown in Fig.~\ref{LAA/Wifi_Queuing}.

To simplify our description, packets to be transmitted by the LAA small cell user are called LAA packets, and those to be transmitted by the Wi-Fi device are Wi-Fi packets.
The Wi-Fi device follows the CSMA/CA protocol to transmit Wi-Fi packets through the unlicensed channel.
We assume that there are $D$ unlicensed channels.
Let $D_{f}$ be the number of free unlicensed channels (where $0 \leq D_{f} \leq D$).

During the occupancy phase, an unlicensed channel may alternately stay in the ON state (indicating there is no Wi-Fi transmission on the unlicensed channel) or the OFF state (indicating there is ongoing Wi-Fi transmission).
We use a timer (denoted by {\bf timer}) to control the length of the occupancy phase for an unlicensed channel.
Expiration of {\bf timer} implies the end of the occupancy phase.

When an LTE packet arrives, if the unlicensed channel is in the OFF state, it will be put into an FIFO queue (of size $Q$) that is implemented at the LAA small cell.
When an LAA packet arrives at the LAA small cell, let $Q_{f}$ denote the number of free spaces (where $0 \leq Q_{f} \leq Q$) of the FIFO queue.
UFA, UTA, and the buffering mechanism can be implemented at the controller at the LAA small.
The details are given below:


Fig.~\ref{LAA/Wifi_Queuing} shows the queuing model for the LAA and Wi-Fi coexisting system.
The details of UFA, UTA, and the buffering mechanism, which are implemented at the controller in the LAA small cell, are given below:
\begin{itemize}
\item \textbf{Unlicensed Full Allocation (UFA):} The controller sets the unlicensed channel to the ON state until there is no LAA packet queued in the FIFO queue.
In other words, the controller switches the unlicensed channel to the OFF state when the FIFO queue is empty.
To implement the UFA, the LAA small cell senses the unlicensed channel via CCA (see Step 1 in Fig.~\ref{Flow}) and negotiates with the UE for the random access and RRC connection in the channel occupancy phase (see Step 2 to Step 6 in Fig.~\ref{Flow}).
After Step 6 in Fig.~\ref{Flow}, the LAA small cell starts to transmit data packets to the UE (see Step 7 in Fig.~\ref{Flow}).
When the LAA small cell completes the transmission, it indicates the last data packet (see Step 12 in Fig.~\ref{Flow}).
Then, the LAA small cell terminates the downlink transmission by the RRC connection Release (see Step 13 in Fig.~\ref{Flow}).

\item \textbf{Unlicensed Time-division Allocation (UTA):} If the FIFO queue is not empty, the controller will execute the sensing phase and the occupancy phase alternatively. The length of the sensing phase and occupancy phase is determined by the {\bf timer} of the sensing phase and occupancy phase, respectively. If the unlicensed server is free, the controller sets the occupancy phase to the ON state. Otherwise, the controller sets the occupancy phase to the OFF state.
For the UTA, the CCA, random access, and RRC connections are the same as the UFA (see Steps 1 to 6 in Fig.~\ref{Flow}).
The difference is that the channel occupancy phase of the UTA has the {\bf timer} to limit the length of the channel occupancy phase (see Step 7 in Fig.~\ref{Flow}).
If the LAA small cell requires more data transmission time, it will re-execute the CCA for the next channel occupancy phase (see
Step 8 in Fig.~\ref{Flow}) and reallocate the resources for downlink transmission (see Step 9 to Step 11 in Fig.~\ref{Flow}).
When the LAA small cell completes the transmission, the UTA is the same as the UFA (see Step 12 and Step 13 in Fig.~\ref{Flow}).

\item \textbf{Buffer Mechanism:} When an LAA packet arrives, the LAA small cell checks the status of the FIFO queue and the unlicensed channel pool. We consider three cases:
\begin{itemize}
\item \textbf{Case 1 :} $Q_{f} \geq Q_{\theta}$ (where $Q_{\theta}$ is a threshold for buffering control). The LAA packet will stay in the FIFO queue.
\item \textbf{Case 2 :} $Q_{f} < Q_{\theta}$ and $D_{f} > 0$ The LAA small cell assigns one unlicensed channel to serve the LAA packet.
\item \textbf{Case 3 :} $Q_{f} = 0$ and $D_{f} = 0$ The LAA packet is dropped.
\end{itemize}
To implement the buffer mechanism, the LAA small cell will check the buffer before Step 1 and Step 8 in Fig.~\ref{Flow}.
If the number of data packets in the buffer is larger than the preset threshold $Q_{\theta}$, the LAA small cell will execute Step 1 and Step 8.
Otherwise, the data packet will be buffered in the buffer of the LAA small cell.
The buffer mechanism can be combined with UFA or UTA as UFAB and UTAB, respectively.
\end{itemize}


\section{Analytical Models}
\label{sect:analytical}

In this section, we propose analytical models to investigate the performance of UFA and UTA with or without the buffering mechanism.
We also develop simulation models to validate the analytical models.
The simulation approach is similar to that in~\cite{Chou2019,Phone2001,Phone2003}.
The output measures include
the packet dropping probability $P_{b,\ell}$ for the LAA packets, and
the packet dropping probability $P_{b, w}$ for the Wi-Fi packets.

Consider Fig.~\ref{LAA/Wifi_Queuing}.
The periods of the sensing phase of the LAA small cell are assumed to have an exponential distribution with mean $1/\mu_s$.
We suppose that the time period for the occupancy phase has an exponential distribution with the mean $1/\delta$.
The LAA packet arrivals and Wi-Fi packet arrivals to a cell are assumed to form Poisson processes with rates $\lambda_{\ell}$ and $\lambda_{w}$, respectively.
The transmission time of the LAA packet $t_{\ell, u}$ on an unlicensed channel is assumed to have exponential distribution with the density function $f_{\ell, u}(t_{\ell, u})=\mu_{\ell, u}e^{-\mu_{\ell, u}t_{\ell, u}}$ and
$E[t_{\ell, u}] =1/\mu_{\ell, u}$. The Wi-Fi packet transmission time on an unlicensed channel, $t_{w}$, is assumed to have exponential distribution with the density function
$f_{w}(t_{w})=\mu_{w}e^{-\mu_{w}t_{w}}$ and
$E[t_{w}] =1/\mu_{w}$.

Note that in real systems, the inter-arrival and transmission times of LAA or Wi-Fi packets may not follow exponential distributions.
Our analytic models assume exponential distributions for two reasons.
First, the exponential distribution can provide the mean value analysis.
Second, the analytic models are for validation of the simulation experiments (where other distributions can be set easily), based on which we can study the effects of different distributions on the performance.

The behavior of the LAA and Wi-Fi coexisting network can be modeled by a Markov process with the system state $\mathbf{n}= (w, x, y, z)$.
$w$ denotes the status of an unlicensed channel, where $w=1$ implies that the LAA system is performing the CCA (i.e., Step 1 or Step 8 in Fig.~\ref{Flow}), and $w=1$ and $w=2$ imply that the unlicensed channel is at the {OFF} state and the {ON} state, respectively.
To solve the system, we employ the finite Markov chain approximation by limiting the number of unlicensed band servers to one.
$x = 1$ and $x = 0$ imply that there is an LAA packet being served by the unlicensed channel and no LAA packet being served by the unlicensed channel, respectively.
$y = 1$ and $y = 0$ imply that there is a Wi-Fi packet being served by the unlicensed channel and no Wi-Fi packet being served by the unlicensed channel, respectively.
$z$ denotes the number of LAA packets buffered in the FIFO queue. The ON/OFF ``\textbf{timer}" of the channel occupation phase for LAA small cell is also assumed to have exponential distribution with the density function $f_{on}(t_{on})=\mu_{on}e^{-\mu_{on}t_{on}}$ for ON state and $f_{off}(t_{off})=\mu_{off}e^{-\mu_{off}t_{off}}$ for OFF state, where $E[t_{on}] =1/\mu_{on}$ and $E[t_{off}] =1/\mu_{off}$.

\subsection{Analytical Model for the UFA}
\label{sect:analytical_UFA}

\begin{figure}[t]
\centering
\includegraphics[width=.5\textwidth]{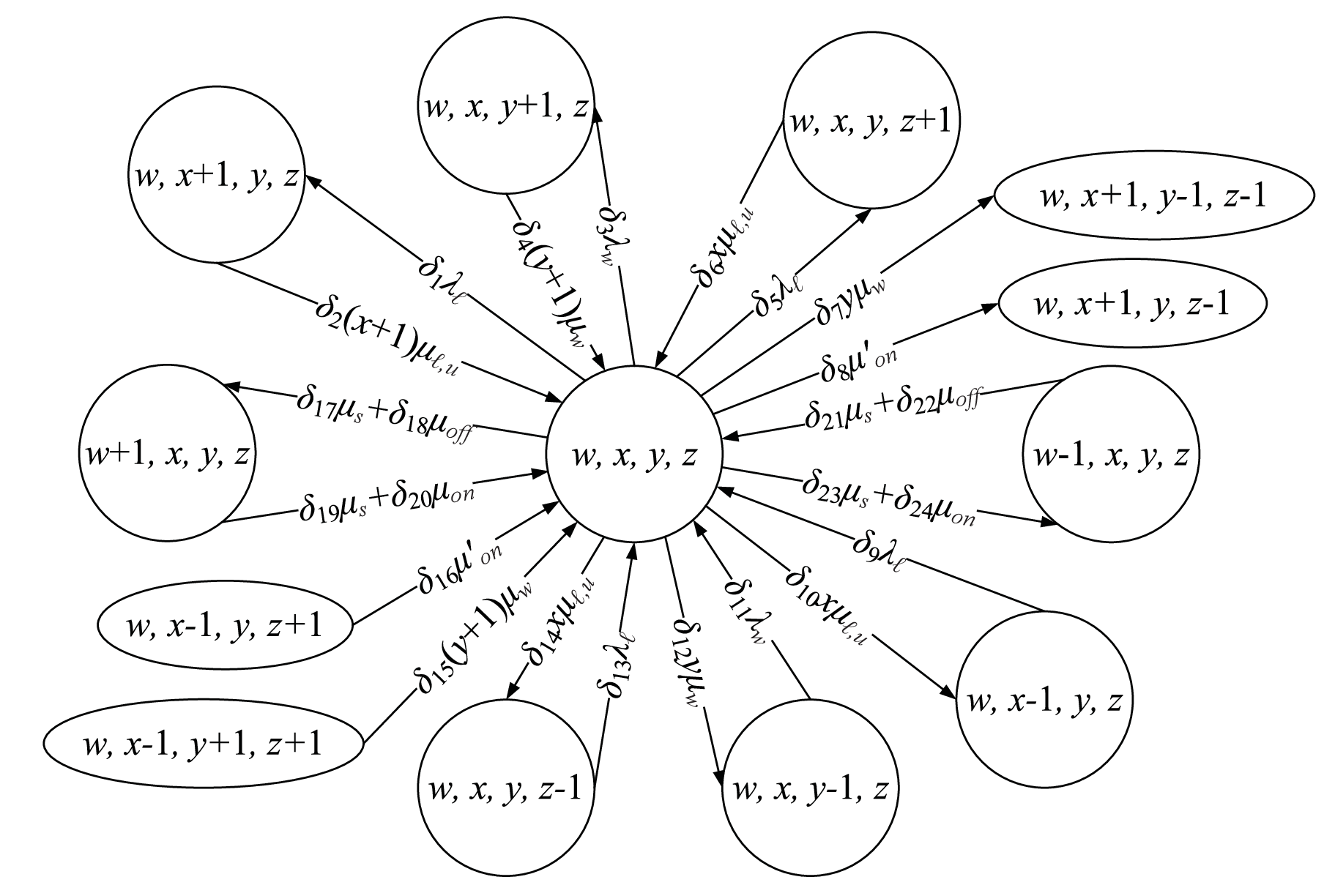}
\caption{The state transition diagram for the UFA, UTA, and buffer mechanism scheme.}
\label{Basic_Queuing}
\end{figure}

The state space of the Markov process for UFA is
\begin{align}
\mathbf{S}_{UFA} = &\big\{\mathbf{n} |  w = 2, x \in \{0, 1\}, y \in \{0, 1\}, z \in \{0, 1, \ldots , Q\} \big\}.
\end{align}
Note that because in UFA, the unlicensed channel is always in the ON state, $w$ is fixed to $2$.
Let $\pi_{w,x,y,z}$ be the steady state probability for state $(w,
x, y, z)$, where $\pi_{w,x,y,z} = 0$ if state $(w,x,y,z) \notin
\mathbf{S}_{UFA}$.
For all legal states $(w,x,y,z)$,
$$\sum_{w,x,y,z \in \mathbf{S}_{UFA}}\pi_{w,x,y,z}= 1$$

Fig.~\ref{Basic_Queuing} illustrates the state transition diagram for UFA.
The transitions of the Markov process are described as follows:
\begin{itemize}
\item[\textbf{1)}] If an LAA packet arrives when the process is at state $(w, x, y, z) \in \mathbf{S}_{UFA}$ where there is a free unlicensed channel and the FIFO queue is empty, then the free unlicensed channel is assigned to this LAA packet.
Therefore, the transition from states $(w, x,
y, z)$ to $(w, x+1, y, z)$ occurs only when $0 \leq x < D-y,
z=0$. Define $\delta_{1}$ as
\begin{align}
\delta_{1} &= \left\{
\begin{array}{ll}
1, & 0 \leq x < D-y, z=0\\
&\textrm{and} \;\; (w, x, y, z) \in \mathbf{S}_{UFA}\\
0, & \textrm{otherwise}.
\end{array}\right.
\label{UFA_1}
\end{align}
The process moves from state $(w, x, y, z)$ to $(w, x+1, y, z)$
with rate $\delta_{1}\lambda_{\ell}$.

\item[\textbf{2)}] If the transmission for an LAA packet on an unlicensed channel completes at state $(w, x+1, y, z) \in
\mathbf{S}_{UFA}$ and the FIFO queue is empty, then the unlicensed band server will be released.
Therefore, the transition from states $(w, x+1, y, z)$ to $(w, x, y, z)$ occurs only when $0 < x+1 \leq D-y, z=0$.
Define $\delta_{2}$ as
\begin{align}
\delta_{2}= \left\{
\begin{array}{ll}
1, & 0 < x+1 \leq D-y, z=0 \\
&\textrm{and} \;\; (w, x+1, y, z) \in \mathbf{S}_{UFA}\\
0, & \textrm{otherwise}.
\end{array}\right.
\label{UFA_2}
\end{align}
The process moves from state $(w, x+1, y, z)$ to $(w, x, y, z)$ with rate $\delta_{2}(x+1)\mu_{\ell, u}$.

\item[\textbf{3)}] If a Wi-Fi packet arrives when the process is at state $(w, x, y, z) \in \mathbf{S}_{UFA}$ and the queue is empty, then one unlicensed band server is
assigned to this Wi-Fi packet. Therefore, the transition from
states $(w, x, y, z)$ to $(w, x, y+1, z)$ occurs only when $0
\leq y < D-x, z=0$. Define $\delta_{3}$ as
\begin{align}
\delta_{3}= \left\{
\begin{array}{ll}
1, & 0 \leq y < D-x, z=0\\
&\textrm{and} \;\; (w, x, y, z) \in \mathbf{S}_{UFA}\\
0, & \textrm{otherwise}.
\end{array}\right.
\label{UFA_3}
\end{align}
The process moves from state $(w, x, y, z)$ to $(w, x, y+1, z)$
with rate $\delta_{3}\lambda_{w}$.

\item[\textbf{4)}] If the transmission for a Wi-Fi packet, which one allocated unlicensed band server completes at state $(w, x, y+1, z) \in
\mathbf{S}_{UFA}$ and the queue is empty, then one unlicensed band server will be released. Therefore, the
transition from states $(w, x, y+1, z)$ to $(w, x, y, z)$ occurs
only when $0 < y+1 \leq D-x, z=0$. Define $\delta_{4}$ as
\begin{align}
\delta_{4}= \left\{
\begin{array}{ll}
1, & 0 < y+1 \leq D-x, z=0\\
&\textrm{and} \;\; (w, x, y+1, z) \in \mathbf{S}_{UFA}\\
0, & \textrm{otherwise}.
\end{array}\right.
\end{align}
The process moves from state $(w, x, y+1, z)$ to $(w, x, y, z)$
with rate $\delta_{4}(y+1)\mu_{w}$.

\item[\textbf{5)}] If an LAA packet arrives at state $(w, x, y, z) \in \mathbf{S}_{UFA}$ where there is no free unlicensed channel, and the queue is not full, then this LAA packet
is buffered in the queue. Therefore, the transition from
states $(w, x, y, z)$ to $(w, x, y, z+1)$ occurs only when $0
\leq z < Q, x+y = D$. Define $\delta_{5}$ as
\begin{align}
\delta_{5}= \left\{
\begin{array}{ll}
1, & 0 \leq z < Q, x+y = D\\
&\textrm{and} \;\; (w, x, y, z) \in \mathbf{S}_{UFA}\\
0, & \textrm{otherwise}.
\end{array}\right.
\end{align}
The process moves from state $(w, x, y, z)$ to $(w, x, y, z+1)$
with rate $\delta_{5}\lambda_{\ell}$.

\item[\textbf{6)}] If the transmission for an LAA packet, which one allocated unlicensed band completes at state $(w, x, y, z+1) \in
\mathbf{S}_{UFA}$, then one LAA packet in the queue will
be served. Therefore, the transition from states $(w, x, y,
z+1)$ to $(w, x, y, z)$ occurs only when $0 < z+1 \leq Q, x+y = D$. Define $\delta_{6}$ as
\begin{align}
\delta_{6}= \left\{
\begin{array}{ll}
1, & 0 < z+1 \leq Q, x+y = D, x > 0\\
&\textrm{and} \;\; (w, x, y, z+1) \in \mathbf{S}_{UFA}\\
0, & \textrm{otherwise}.
\end{array}\right.
\end{align}
The process moves from state $(w, x, y, z+1)$ to $(w, x, y, z)$
with rate $\delta_{6}x\mu_{\ell, u}$.

\item[\textbf{7)}] If the transmission for a Wi-Fi packet, which one allocated unlicensed band server completes at state $(w, x, y, z) \in
\mathbf{S}_{UFA}$, then one LAA packet in the queue will
be served. Therefore, the transition from states $(w, x, y, z)$
to $(w, x+1, y-1, z-1)$ occurs only when $0 < y, 0 < z, x+y = D$. Define $\delta_{7}$ as
\begin{align}
\delta_{7}= \left\{
\begin{array}{ll}
1, & y > 0, z > 0, x+y = D\\
&\textrm{and} \;\; (w, x, y, z) \in \mathbf{S}_{UFA}\\
0, & \textrm{otherwise}.
\end{array}\right.
\label{UFA_7}
\end{align}
The process moves from state $(w, x, y, z)$ to $(w, x+1, y-1, z-1)$
with rate $\delta_{7}y\mu_{w}$.
\end{itemize}

The transitions between $(w, x, y, z)$ and $(w, x-1, y, z)$, $(w, x, y-1, z)$, $(w, x, y, z-1)$, $(w, x-1, y+1,
z+1)$ are similar to that between $(w, x, y, z)$ and $(w+1, x, y,
z)$, $(w, x+1, y, z)$, $(w, x, y+1, z)$, $(w, x, y, z+1)$, $(w, x+1,
y-1, z-1)$.

The Markov process has the \emph{reversible property} if it satisfies the balance equations.
The balance equations require that the transition probability matrix, $P$, for the Markov process possess a steady state probability $\pi$ such that $\pi_{i}P_{i,j} = \pi_{j}P_{j, i}$, where $P_{i,j}$ is the Markov transition probability (i.e., transition rate) from state $i$ to state $j$, i.e. $P_{i,j} = \textrm{Prob}(X_{t} = j | X_{t-1}= i)$, and $\pi_{i}$ and $\pi_{j}$ are the steady state probabilities of being in states $i$ and $j$, respectively.

For this process, we can formulate the balance equations as the steady state probability of the current state $(w, x, y, z)$ multiple by the sum of the transition rates which switch to the other states equal to the steady state probabilities of the other states multiplied by the transition rates at which transitions to the current state $(w, x, y, z)$. Therefore, the balance equations for this process are:
\begin{align}
\pi_{w, x, y, z} = \frac{\varepsilon_{A}}{\varepsilon_{B}},
\label{Balance Equation}
\end{align}
where
\begin{align}
\varepsilon_{A} = &\; \delta_{2}(x+1)\mu_{\ell, u}\pi_{w, x+1, y, z} + \delta_{4}(y+1)\mu_{w}\pi_{w, x, y+1, z} \notag\\
&+ \delta_{6}x\mu_{\ell, u}\pi_{w, x, y, z+1}+ \delta_{9}\lambda_{\ell}\pi_{w, x-1, y, z} + \delta_{11}\lambda_{w}\pi_{w, x, y-1, z}\notag\\
&+ \delta_{13}\lambda_{\ell}\pi_{w, x, y, z-1}+ \delta_{15}(y+1)\mu_{w}\pi_{w, x-1, y+1, z+1} \notag\\
&+ \delta_{16}\mu_{on}'\pi_{w, x-1, y, z+1}+ (\delta_{19}\mu_{s} + \delta_{20}\mu_{on})\pi_{w+1, x, y, z}\notag\\
&+ (\delta_{21}\mu_{s} + \delta_{22}\mu_{off})\pi_{w-1, x, y, z}.
\end{align}
and
\begin{align}
\varepsilon_{B} = &\; \delta_{1}\lambda_{\ell} + \delta_{3}\lambda_{w} + \delta_{5}\lambda_{\ell} + \delta_{7}y\mu_{w} + \delta_{8}\mu_{on}' + \delta_{10}x\mu_{\ell, u}\notag\\
&+ \delta_{12}y\mu_{w} + \delta_{14}x\mu_{\ell, u} + \delta_{17}\mu_{s} + \delta_{18}\mu_{off} + \delta_{23}\mu_{s}\notag\\
&+ \delta_{24}\mu_{on},
\end{align}

Note that $\delta_{1}$, $\delta_{2}$, \ldots , $\delta_{7}$ are
obtained from~(\ref{UFA_1}),~(\ref{UFA_2}),~(\ref{UFA_3}), \ldots ,~(\ref{UFA_7}), respectively, and $\delta_{8} = \delta_{16} = \delta_{17} = \delta_{18} = \delta_{19} = \delta_{20} = \delta_{21} = \delta_{22} = \delta_{23} = \delta_{24} = 0$,
\begin{align}
\delta_{9}&= \left\{
\begin{array}{ll}
1, & 0 \leq x-1 < D-y, z=0\\
&\textrm{and} \;\; (w, x-1, y, z) \in \mathbf{S}_{UFA}\\
0, & \textrm{otherwise}.
\end{array}\right.\\
\delta_{10}&= \left\{
\begin{array}{ll}
1, & 0 < x \leq D-y, z=0\\
&\textrm{and} \;\; (w, x, y, z) \in \mathbf{S}_{UFA}\\
0, & \textrm{otherwise}.
\end{array}\right.\\
\delta_{11}&= \left\{
\begin{array}{ll}
1, & 0 \leq y-1 < D-x, z=0\\
&\textrm{and} \;\; (w, x, y-1, z) \in \mathbf{S}_{UFA}\\
0, & \textrm{otherwise}.
\end{array}\right.\\
\delta_{12}&= \left\{
\begin{array}{ll}
1, & 0 < y \leq D-x, z=0\\
&\textrm{and} \;\; (w, x, y, z) \in \mathbf{S}_{UFA}\\
0, & \textrm{otherwise}.
\end{array}\right.\\
\delta_{13}&= \left\{
\begin{array}{ll}
1, & 0 \leq z-1 < Q, x+y = D,\\
&\textrm{and} \;\; (w, x, y, z-1) \in \mathbf{S}_{UFA}\\
0, & \textrm{otherwise}.
\end{array}\right.\\
\delta_{14}&= \left\{
\begin{array}{ll}
1, & 0 < z \leq Q, x+y = D, x > 0\\
&\textrm{and} \;\; (w, x, y, z) \in \mathbf{S}_{UFA}\\
0, & \textrm{otherwise}.
\end{array}\right.\\
\delta_{15}&= \left\{
\begin{array}{ll}
1, & 0 < y+1, 0 < z+1, x+y = D\\
&\textrm{and} \;\; (w, x-1, y+1, z+1) \in \mathbf{S}_{UFA}\\
0, & \textrm{otherwise}.
\end{array}\right.
\label{UFA_9-15}
\end{align}

When an LAA packet arrives at the states where the queue is
full, and there is no free unlicensed server (i.e.,
$D_{f}=0, Q_{f}=0$), this LAA packet will be dropped. On
the other hand, when a Wi-Fi packet arrives at the states where
there is no free unlicensed band server (i.e., $D_{f} = 0$) and the unlicensed band serving packet is LAA (i.e., $x \neq 0$), this Wi-Fi packet will be dropped. Therefore,
\begin{align}
\label{block_prob_LAA}
P_{b, \ell} &= \sum_{(w, x, y, z) \in \{\mathbf{n} | z = Q, \mathbf{n} \in \mathbf{S}_{UFA}\}}\pi_{w, x, y, z},\\
\label{block_prob_WF}
P_{b, w} &= \sum_{(w, x, y, z) \in \{\mathbf{n} | x = D, \mathbf{n} \in \mathbf{S}_{UFA}\}}\pi_{w, x, y, z}.
\end{align}

From (\ref{Balance Equation}), (\ref{block_prob_LAA}), and
(\ref{block_prob_WF}), the steady state probabilities $\pi_{w, x, y,
z}$ and $P_{b, \ell}$ and $P_{b, w}$ can be computed by the
iterative algorithm in~\cite{Chou2019,Phone2001,Phone2003} for LAA coexisting with
Wi-Fi heterogeneous networks as
\begin{itemize}
\item \textbf{Step 1.}  Select an initial value for $\pi(\mathbf{n})$.
\item \textbf{Step 2.} $\pi_{\textrm{old}}(\mathbf{n}) \leftarrow \pi(\mathbf{n})$.
\item \textbf{Step 3.} Compute $\pi(\mathbf{n})$ by using (\ref{Balance Equation}).
\item \textbf{Step 4.} Compute $G^{-1}\pi(\mathbf{n})$. The normalized factor $G=\sum_{\mathbf{n} \in \mathbf{S}} \pi(\mathbf{n})$ is used to ensure that $\sum_{\mathbf{n} \in \mathbf{S}} \pi(\mathbf{n}) = 1$.
\item \textbf{Step 5.} If $|\pi(\mathbf{n}) - \pi_{\textrm{old}}(\mathbf{n})| > \alpha \pi(\mathbf{n}) $ then go to \textbf{Step 2.)}. Otherwise, go to \textbf{Step 6.}. Note that $\alpha$ is a predefined $10^{-6}$
\item \textbf{Step 6.} The stationary probability $\pi(\mathbf{n})$ of the state $\mathbf{n}= (w, x, y, z) \in \mathbf{S}_{UFA}$ converge. Compute $P_{b, \ell}$, $P_{b, w}$ by using (\ref{block_prob_LAA}) and (\ref{block_prob_WF}).
\end{itemize}

In all cases considered in this paper, the above iterative algorithm always
converges. The simulation experiments indicate that the iterative algorithm
converges to the correct values (see Section~\ref{sect:validation}).

\subsection{Analytical Model for the UTA}
\label{sect:analytical_UTA}

This section proposes an analytic model for UTA. We also
use the Zachary-Kelly model~\cite{Zachary1991,Kelly1991} together
with an iterative algorithm to derive the LAA and Wi-Fi packet
dropping probability (i.e., $P_{b, \ell}$ and $P_{b, w}$) in the section~\ref{sect:analytical_UFA}.

Based on the description of the UFA in the previous section, the network
assigns channels to every packet in the coexistence system with the
UTA, the state space
$\mathbf{S}_{UTA}$ can be expressed as
\begin{align}
\mathbf{S}_{UTA} = \big\{&\mathbf{n} | w \in \{0, 1, 2\}, x \in \{0, 1\}, y \in \{0, 1\},\notag \\
&z \in \{0, 1, \ldots , Q\} \big\}.
\end{align}

In addition, we let $\pi_{w,x,y,z}$ be the steady state probability
for state $(w, x, y, z)$, where $\pi_{w,x,y,z} = 0$ if state
$(w,x,y,z) \notin \mathbf{S}_{UTA}$. For all legal states $(w,x,y,z)
\in \mathbf{S}_{UTA}$, $\sum_{w,x,y,z \in
\mathbf{S}_{UTA}}\pi_{w,x,y,z} = 1$.

The state transition diagram for an LAA coexisting with Wi-Fi
heterogeneous network is illustrated as~Fig.~\ref{Basic_Queuing} in
the previous section. The transitions of the Markov process are
described as follows. If state $(w, x, y, z) \in \mathbf{S}_{UTA}$,
the following transitions should be considered.

The balance equations for the Markov process are also expressed
as~(\ref{Balance Equation}), and $\delta_{1}$, $\delta_{2}$, \ldots
, $\delta_{18}$ are obtained in the following equations:

\begin{itemize}
\item[\textbf{1)}] If an LAA packet arrives when the
process is at state $(w, x, y, z) \in \mathbf{S}_{UTA}$ where the state of the LAA small cell is ON and the queue is
empty, then one unlicensed band server is assigned to this LAA
packet. Therefore, the transition from states $(w, x, y, z)$ to
$(w, x+1, y, z)$ occurs only when $w=2, 0 \leq x < D-y, z=0$.
Define $\delta_{1}$ as
\begin{align}
\delta_{1}= \left\{
\begin{array}{ll}
1, & w=2, 0 \leq x < D-y, z=0\\
&\textrm{and} \;\; (w, x, y, z) \in \mathbf{S}_{UTA}\\
0, & \textrm{otherwise}.
\end{array}\right.
\label{UTA_1}
\end{align}
The process moves from state $(w, x, y, z)$ to $(w, x+1, y, z)$
with rate $\delta_{1}\lambda_{\ell}$.

\item[\textbf{2)}] If the transmission for an LAA packet, which one allocated unlicensed band server completes at state $(w, x+1, y, z) \in
\mathbf{S}_{UTA}$, then one unlicensed band server will be released. Therefore, the
transition from states $(w, x+1, y, z)$ to $(w, x, y, z)$ occurs when $w=2, 0 < x \leq D-y, z=0$ or $w \neq 2, 0 < x \leq D-y$. Define $\delta_{2}$ as
\begin{align}
\delta_{2}= \left\{
\begin{array}{ll}
1, & (w=2, 0 < x \leq D-y, z=0)\\
&\cup (w \neq 2, 0 < x \leq D-y)\\
&\textrm{and} \;\; (w, x+1, y, z) \in \mathbf{S}_{UTA}\\
0, & \textrm{otherwise}.
\end{array}\right.
\label{UTA_2}
\end{align}
The process moves from state $(w, x+1, y, z)$ to $(w, x, y, z)$
with rate $\delta_{2}(x+1)\mu_{\ell, u}$.

\item[\textbf{3)}] If a Wi-Fi packet arrives when the
process is at state $(w, x, y, z) \in \mathbf{S}_{UTA}$, then one unlicensed band server is
assigned to this Wi-Fi packet. Therefore, the transition from
states $(w, x, y, z)$ to $(w, x, y+1, z)$ occurs only when $0
\leq y < D-x$. Define $\delta_{3}$ as
\begin{align}
\delta_{3}= \left\{
\begin{array}{ll}
1, & 0 \leq y < D-x\\
&\textrm{and} \;\; (w, x, y, z) \in \mathbf{S}_{UTA}\\
0, & \textrm{otherwise}.
\end{array}\right.
\label{UTA_3}
\end{align}
The process moves from state $(w, x, y, z)$ to $(w, x, y+1, z)$
with rate $\delta_{3}\lambda_{w}$.

\item[\textbf{4)}] If the transmission for a Wi-Fi packet, which one allocated unlicensed band server completes at state $(w, x, y+1, z) \in
\mathbf{S}_{UTA}$, then one unlicensed band server will be released. Therefore, the
transition from states $(w, x, y+1, z)$ to $(w, x, y, z)$ occurs
only when $0 < y+1 \leq D-x$. Define $\delta_{4}$ as
\begin{align}
\delta_{4}= \left\{
\begin{array}{ll}
1, & 0 < y+1 \leq D-x\\
&\textrm{and} \;\; (w, x, y+1, z) \in \mathbf{S}_{UTA}\\
0, & \textrm{otherwise}.
\end{array}\right.
\end{align}
The process moves from state $(w, x, y+1, z)$ to $(w, x, y, z)$
with rate $\delta_{4}(y+1)\mu_{w}$.

\item[\textbf{5)}] If an LAA packet arrives at state $(w, x, y, z) \in
\mathbf{S}_{UTA}$ where there is no free unlicensed
channel or the state of the LAA small cell is not ON, and the queue is not full, then this LAA packet
is buffered in the queue. Therefore, the transition from
states $(w, x, y, z)$ to $(w, x, y, z+1)$ occurs only when $0
\leq z < Q, x+y = D$. Define $\delta_{5}$ as
\begin{align}
\delta_{5}= \left\{
\begin{array}{ll}
1, & (w=2, 0 \leq z < Q, x+y=D)\\
&\cup (w \neq 2, 0 \leq z < Q)\\
&\textrm{and} \;\; (w, x, y, z) \in \mathbf{S}_{UTA}\\
0, & \textrm{otherwise}.
\end{array}\right.
\end{align}
The process moves from state $(w, x, y, z)$ to $(w, x, y, z+1)$
with rate $\delta_{5}\lambda_{\ell}$.

\item[\textbf{6)}] If the transmission for an LAA packet, which allocated one unlicensed band serve, completes at state $(w, x, y, z+1) \in
\mathbf{S}_{UTA}$ and the state of the LAA small cell is ON, then one LAA packet in the queue will
be served. Therefore, the transition from states $(w, x, y,
z+1)$ to $(w, x, y, z)$ occurs only when $w=2, 0 < z+1 \leq Q,
x+y = D$. Define $\delta_{6}$ as
\begin{align}
\delta_{6}= \left\{
\begin{array}{ll}
1, & w=2, 0 < z+1 \leq Q, x+y = D\\
&\textrm{and} \;\; (w, x, y, z+1) \in \mathbf{S}_{UTA}\\
0, & \textrm{otherwise}.
\end{array}\right.
\end{align}
The process moves from state $(w, x, y, z+1)$ to $(w, x, y, z)$
with rate $\delta_{6}x\mu_{\ell, u}$.

\item[\textbf{7)}] If the transmission for a Wi-Fi packet, which one allocated unlicensed band server completes at state $(w, x, y, z) \in
\mathbf{S}_{UTA}$ and the state of LAA small cell is ON, then one LAA packet in the queue will
be served. Therefore, the transition from states $(w, x, y, z)$
to $(w, x+1, y-1, z-1)$ occurs only when $w=2, y > 0, z > 0, x+y = D$. Define $\delta_{7}$ as
\begin{align}
\delta_{7}= \left\{
\begin{array}{ll}
1, & w=2, y > 0, z > 0, x+y = D\\
&\textrm{and} \;\; (w, x, y, z) \in \mathbf{S}_{UTA}\\
0, & \textrm{otherwise}.
\end{array}\right.
\end{align}
The process moves from state $(w, x, y, z)$ to $(w, x+1, y-1,
z-1)$ with rate $\delta_{7}y\mu_{w}$.

\item[\textbf{8)}] When the unlicensed band server is free, if the state of the LAA small cell is switched to ON, the transmission for a LAA packet that one allocated unlicensed band server completes at state $(w, x, y, z) \in \mathbf{S}_{UTA}$ and the queue is not empty, then one LAA packet in the queue will
be served. Therefore, the transition from states $(w, x, y, z)$
to $(w, x+1, y, z-1)$ occurs only when $w=2, y = 0, z > 0, x+y = D$. Define $\delta_{8}$ as
\begin{align}
\delta_{8}= \left\{
\begin{array}{ll}
1, & w = 2, y = 0, z > 0, 0 \leq x < D-y\\
&\textrm{and} \;\; (w, x, y, z) \in \mathbf{S}_{UTA}\\
0, & \textrm{otherwise}.
\end{array}\right.
\label{UTA_8}
\end{align}
The process moves from state $(w, x, y, z)$ to $(w, x+1, y,
z-1)$ with rate $\delta_{8}\mu_{on}'$, where $\mu_{on}' = 10\mu_{on}$.

The transitions between $(w, x, y, z)$ and $(w, x-1, y, z)$, $(w, x, y-1, z)$, $(w, x, y, z-1)$, $(w, x-1,
y+1, z+1)$, $(w, x-1, y, z+1)$ are similar to that between $(w, x, y, z)$ and $(w+1,
x, y, z)$, $(w, x+1, y, z)$, $(w, x, y+1, z)$, $(w, x, y, z+1)$,
$(w, x+1, y-1, z-1)$, $(w, x+1, y, z-1)$. The balance equations for this process
is~(\ref{Balance Equation}), where $\delta_{1}$, $\delta_{2}$,
\ldots , $\delta_{8}$ are obtained
from~(\ref{UTA_1}),~(\ref{UTA_2}),~(\ref{UTA_3}), \ldots
,~(\ref{UTA_8}), respectively, and

\begin{align}
\delta_{9} &= \left\{
\begin{array}{ll}
1, & w=2, 0 \leq x-1 < D-y, z=0\\
&\textrm{and} \;\; (w, x-1, y, z) \in \mathbf{S}_{UTA}\\
0, & \textrm{otherwise}.
\end{array}\right.\\
\delta_{10} &= \left\{
\begin{array}{ll}
1, & (w=2, 0 < x \leq D-y, z=0)\\
&\cup (w \neq 2, 0 < x \leq D-y)\\
&\textrm{and} \;\; (w, x, y, z) \in \mathbf{S}_{UTA}\\
0, & \textrm{otherwise}.
\end{array}\right.\\
\delta_{11} &= \left\{
\begin{array}{ll}
1, & 0 \leq y-1 < D-x\\
&\textrm{and} \;\; (w, x, y-1, z) \in \mathbf{S}_{UTA}\\
0, & \textrm{otherwise}.
\end{array}\right.
\end{align}
\begin{align}
\delta_{12} &= \left\{
\begin{array}{ll}
1, & 0 \leq y < D-x\\
&\textrm{and} \;\; (w, x, y, z) \in \mathbf{S}_{UTA}\\
0, & \textrm{otherwise}.
\end{array}\right.\\
\delta_{13} &= \left\{
\begin{array}{ll}
1, & (w=2, 0 \leq z-1 < Q, x+y=D)\\
&\cup (w \neq 2, 0 \leq z-1 < Q)\\
&\textrm{and} \;\; (w, x, y, z-1) \in \mathbf{S}_{UTA}\\
0, & \textrm{otherwise}.
\end{array}\right.\\
\delta_{14} &= \left\{
\begin{array}{ll}
1, & w=2, 0 < z \leq Q\\
&\textrm{and} \;\; (w, x, y, z) \in \mathbf{S}_{UTA}\\
0, & \textrm{otherwise}.
\end{array}\right.\\
\delta_{15} &= \left\{
\begin{array}{ll}
1, & w=2, y+1>0, z+1>0, x+y=D\\
&\textrm{and} \;\; (w, x-1, y+1, z+1) \in \mathbf{S}_{UTA}\\
0, & \textrm{otherwise}.
\end{array}\right.\\
\delta_{16} &= \left\{
\begin{array}{ll}
1, & w=2, y=0, z+1>0,\\
&0 \leq x-1 < D-y\\
&\textrm{and} \;\; (w, x-1, y, z+1) \in \mathbf{S}_{UTA}\\
0, & \textrm{otherwise}.
\end{array}\right.
\label{UTA_9-16}
\end{align}
\end{itemize}

The state transitions of the LAA small cell ($w \in \{0 (\textrm{OFF}), 1 (\textrm{Sensing}), 2 (\textrm{ON})\}$) are described as follows, and $\delta_{17}$, $\delta_{18}$, \ldots
, $\delta_{24}$ are obtained in the following equations:

\begin{itemize}
\item[\textbf{9)}] When the state of the LAA small cell is Sensing, if the unlicensed band server is free and the queue is not empty, then the state of the LAA small cell is switched to ON. Therefore, the transition from states $(w, x, y, z)$ to $(w+1, x, y, z)$ occurs only when $w=1, x=0, y=0, z>0$. Define $\delta_{17}$ as
\begin{align}
\delta_{17} &= \left\{
\begin{array}{ll}
1, & w=1, x=0, y=0, z>0\\
&\textrm{and} \;\; (w, x, y, z) \in \mathbf{S}_{UTA}\\
0, & \textrm{otherwise}.
\end{array}\right.
\label{UTA_17}
\end{align}
The process moves from state $(w, x, y, z)$ to $(w+1, x, y,
z)$ with rate $\delta_{17}\mu_{s}$.

\item[\textbf{10)}] When the state of the LAA small cell is OFF, if the queue is not empty, then the state of the LAA small cell is switched to Sensing. Therefore, the transition from states $(w, x, y, z)$ to $(w+1, x, y, z)$ occurs only when $w=0, z>0$. Define $\delta_{18}$ as
\begin{align}
\delta_{18} &= \left\{
\begin{array}{ll}
1, & w=0, z>0\\
&\textrm{and} \;\; (w, x, y, z) \in \mathbf{S}_{UTA}\\
0, & \textrm{otherwise}.
\end{array}\right.
\label{UTA_18}
\end{align}
The process moves from state $(w, x, y, z)$ to $(w+1, x, y,
z)$ with rate $\delta_{18}\mu_{off}$.

\item[\textbf{11)}] When the state of the LAA small cell is Sensing, if the queue is occupied by a Wi-Fi packet or the queue is empty, then the state of the LAA small cell is switched to OFF. Therefore, the transition from states $(w+1, x, y, z)$ to $(w, x, y, z)$ occurs only when $w+1=1, x=0, y=1$ or $w+1=1, z=0$. Define $\delta_{19}$ as
\begin{align}
\delta_{19} &= \left\{
\begin{array}{ll}
1, & (w+1=1, x=0, y=1)\\
&\cup (w+1=1, z=0)\\
&\textrm{and} \;\; (w+1, x, y, z) \in \mathbf{S}_{UTA}\\
0, & \textrm{otherwise}.
\end{array}\right.
\end{align}
The process moves from state $(w+1, x, y, z)$ to $(w, x, y,
z)$ with rate $\delta_{19}\mu_{s}$.

\item[\textbf{12)}] When the state of LAA small cell is ON, then the state of LAA small cell is switched to Sensing. Therefore, the transition from states $(w+1, x, y, z)$ to $(w, x, y, z)$ occurs only when $w+1=2$. Define $\delta_{20}$ as
\begin{align}
\delta_{20} &= \left\{
\begin{array}{ll}
1, & w+1=2\\
&\textrm{and} \;\; (w+1, x, y, z) \in \mathbf{S}_{UTA}\\
0, & \textrm{otherwise}.
\end{array}\right.
\end{align}
The process moves from state $(w+1, x, y, z)$ to $(w, x, y,
z)$ with rate $\delta_{20}\mu_{on}$.
\end{itemize}

The transitions between $(w, x, y, z)$ and $(w-1, x, y, z)$ is similar to that between $(w, x, y, z)$ and $(w+1,
x, y, z)$, and $\delta_{21}$, $\delta_{22}$,
\ldots , $\delta_{24}$ are obtained from~(\ref{UTA_21}),~(\ref{UTA_22}), \ldots ,~(\ref{UTA_21-24}), respectively, and
\begin{align}
\delta_{21} &= \left\{
\begin{array}{ll}
1, & w-1=1, x=0, y=0, z>0\\
&\textrm{and} \;\; (w-1, x, y, z) \in \mathbf{S}_{UTA}\\
0, & \textrm{otherwise}.
\end{array}\right.\label{UTA_21}\\
\delta_{22} &= \left\{
\begin{array}{ll}
1, & w-1=0, z>0\\
&\textrm{and} \;\; (w-1, x, y, z) \in \mathbf{S}_{UTA}\\
0, & \textrm{otherwise}.
\end{array}\right.\label{UTA_22}\\
\delta_{23} &= \left\{
\begin{array}{ll}
1, & (w=1, x=0, y=1) \cup (w=1, z=0)\\
&\textrm{and} \;\; (w, x, y, z) \in \mathbf{S}_{UTA}\\
0, & \textrm{otherwise}.
\end{array}\right.\\
\delta_{24} &= \left\{
\begin{array}{ll}
1, & w=2\\
&\textrm{and} \;\; (w, x, y, z) \in \mathbf{S}_{UTA}\\
0, & \textrm{otherwise}.
\end{array}\right.
\label{UTA_21-24}
\end{align}

\subsection{Analytical Model for the UFAB}

This section proposes an analytic model for the UFA combined with the buffering mechanism (UFAB) for LAA packet transmission and significantly
improves the performance of the network. However, the buffering mechanism for LAA packets is not addressed in the previous studies~\cite{SmallCell2012,Liu2011,Liu2015}.
In the UFAB analytic model, the FIFO queues are maintained in the LAA small cell to buffer the LAA packet requests that are not
served immediately due to the fact that there is no free channel or the number of the LAA packets in the FIFO queue is less than the predefined buffering threshold $Q_{\theta}$ (i.e., $z < Q_{\theta}$).
When the number of the LAA packets in the FIFO queue is larger than the buffering threshold $Q_{\theta}$ (i.e., $z \geq Q_{\theta}$), the LAA small cell will allocate unlicensed channels to the LAA packets.
We also
use the Zachary-Kelly model~\cite{Zachary1991,Kelly1991} together
with an iterative algorithm to derive the LAA and Wi-Fi packet
dropping probability (i.e., $P_{b, \ell}$ and $P_{b, w}$) for the UFAB in the
previous section.

Based on the description of the UFA in the section~\ref{sect:analytical_UFA}, the network assigns an unlicensed channel to every LAA packet in the coexistence system with the UFAB. The following constraints must be satisfied, and it is clear that the state space for this Markov process is
\begin{align}
\mathbf{S}_{UFAB} = \big\{&\mathbf{n} | w = 2, x \in \{0, 1\}, y \in \{0, 1\},\notag\\
&z \in \{0, 1, \ldots , Q\}, Q_{\theta} \in \{1, \ldots , Q-1\} \big\}.
\end{align}
The balance equations for the Markov process are also expressed
as~(\ref{Balance Equation}).
Since $\delta_{8}$, $\delta_{16}$, $\delta_{17}$, $\delta_{18}$, $\delta_{19}$, $\delta_{20}$, $\delta_{21}$, $\delta_{22}$, $\delta_{23}$ and $\delta_{24}$ are used to indicate the change of ON/OFF/Sensing states, the UFAB is always at the ON state, so we let $\delta_{8}$, $\delta_{16}$, $\delta_{17}$, $\delta_{18}$, $\delta_{19}$, $\delta_{20}$, $\delta_{21}$, $\delta_{22}$, $\delta_{23}$ and $\delta_{24}$ are equal to zero. The other different Markov conditions are:

\begin{align}
\delta_{1} &= \left\{
\begin{array}{ll}
1, & 0 \leq x < D-y, z \geq Q_{\theta}\\
&\textrm{and} \;\; (w, x, y, z) \in \mathbf{S}_{UFAB}\\
0, & \textrm{otherwise}.
\end{array}\right.\\
\delta_{2}&= \left\{
\begin{array}{ll}
1, & 0 < x+1 \leq D-y, z < Q_{\theta}\\
&\textrm{and} \;\; (w, x+1, y, z) \in \mathbf{S}_{UFAB}\\
0, & \textrm{otherwise}.
\end{array}\right.\\
\delta_{3}&= \left\{
\begin{array}{ll}
1, & 0 \leq y < D-x, z < Q_{\theta}\\
&\textrm{and} \;\; (w, x, y, z) \in \mathbf{S}_{UFAB}\\
0, & \textrm{otherwise}.
\end{array}\right.\\
\delta_{4}&= \left\{
\begin{array}{ll}
1, & 0 < y+1 \leq D-x, z < Q_{\theta}\\
&\textrm{and} \;\; (w, x, y+1, z) \in \mathbf{S}_{UFAB}\\
0, & \textrm{otherwise}.
\end{array}\right.\\
\delta_{5}&= \left\{
\begin{array}{ll}
1, & 0 \leq z < Q, [x+y = D] \cup [z < Q_{\theta}]\\
&\textrm{and} \;\; (w, x, y, z) \in \mathbf{S}_{UFAB}\\
0, & \textrm{otherwise}.
\end{array}\right.\\
\delta_{6}&= \left\{
\begin{array}{ll}
1, & 0 < z+1 \leq Q, x+y = D, x > 0,\\
&z + 1 \geq Q_{\theta} \;\; \textrm{and} \;\; (w, x, y, z+1) \in \mathbf{S}_{UFAB}\\
0, & \textrm{otherwise}.
\end{array}\right.\\
\delta_{7}&= \left\{
\begin{array}{ll}
1, & y > 0, z > 0, x+y = D, z \geq Q_{\theta}\\
&\textrm{and} \;\; (w, x, y, z) \in \mathbf{S}_{UFAB}\\
0, & \textrm{otherwise}.
\end{array}\right.\\
\delta_{9}&= \left\{
\begin{array}{ll}
1, & 0 \leq x-1 < D-y, z \geq Q_{\theta}\\
&\textrm{and} \;\; (w, x-1, y, z) \in \mathbf{S}_{UFAB}\\
0, & \textrm{otherwise}.
\end{array}\right.\\
\delta_{10}&= \left\{
\begin{array}{ll}
1, & 0 < x \leq D-y, z < Q_{\theta}\\
&\textrm{and} \;\; (w, x, y, z) \in \mathbf{S}_{UFAB}\\
0, & \textrm{otherwise}.
\end{array}\right.\\
\delta_{11}&= \left\{
\begin{array}{ll}
1, & 0 \leq y-1 < D-x, z < Q_{\theta}\\
&\textrm{and} \;\; (w, x, y-1, z) \in \mathbf{S}_{UFAB}\\
0, & \textrm{otherwise}.
\end{array}\right.\\
\delta_{12}&= \left\{
\begin{array}{ll}
1, & 0 < y \leq D-x, z < Q_{\theta}\\
&\textrm{and} \;\; (w, x, y, z) \in \mathbf{S}_{UFAB}\\
0, & \textrm{otherwise}.
\end{array}\right.\\
\delta_{13}&= \left\{
\begin{array}{ll}
1, & 0 \leq z-1 < Q, [x+y = D] \cup [z-1 < Q_{\theta}]\\
&\textrm{and} \;\; (w, x, y, z-1) \in \mathbf{S}_{UFAB}\\
0, & \textrm{otherwise}.
\end{array}\right.\\
\delta_{14}&= \left\{
\begin{array}{ll}
1, & 0 < z \leq Q, x+y = D, x > 0,\\
&z \geq Q_{\theta} \;\; \textrm{and} \;\; (w, x, y, z) \in \mathbf{S}_{UFAB}\\
0, & \textrm{otherwise}.
\end{array}\right.\\
\delta_{15}&= \left\{
\begin{array}{ll}
1, & 0 < y+1, 0 < z+1, x+y = D, (z+1) \geq Q_{\theta}\\
&\textrm{and} \;\; (w, x-1, y+1, z+1) \in \mathbf{S}_{UFAB}\\
0, & \textrm{otherwise}.
\end{array}\right.
\end{align}

\subsection{Analytical Model for the UTAB}

Similarly, we also combine the UTA and the buffering mechanism as the UTAB~\cite{Chou2019}.
This section proposes an analytic model for the UTAB, and we also use the Zachary-Kelly model~\cite{Zachary1991,Kelly1991} together
with an iterative algorithm to derive the LAA and Wi-Fi packet
dropping probability (i.e., $P_{b, \ell}$ and $P_{b, w}$) for the UTAB.

Based on the description of the UTA in the section~\ref{sect:analytical_UTA}, the network
assigns channels to every packet in the coexistence system with the
UTAB, the following constraints must be
satisfied, and it is clear that the state space
For this Markov process, it is
\begin{align}
\mathbf{S}_{UTAB} = \big\{&\mathbf{n} | w \in \{0, 1, 2\}, x \in \{0, 1\}, y \in \{0, 1\},\notag\\
&z \in \{0, 1, \ldots , Q\}, Q_{\theta} \in \{1, \ldots , Q-1\} \big\}.
\end{align}
The balance equations for the Markov process are also expressed
as~(\ref{Balance Equation}), where $\delta_{1}$, $\delta_{2}$, \ldots
, $\delta_{16}$, $\delta_{20}$ and $\delta_{24}$ are the same as the UTA in the section~\ref{sect:analytical_UTA} and $\delta_{17}$, $\delta_{18}$, $\delta_{19}$, $\delta_{21}$, $\delta_{22}$ and $\delta_{23}$ obtained in the following equations:

\begin{itemize}
\item[\textbf{1)}] When the state of the LAA small cell is Sensing, if the unlicensed band server is free and the queue is greater than $Q_{\theta}$, then the state of the LAA small cell is switched to ON. Therefore, the transition from states $(w, x, y, z)$ to $(w+1, x, y, z)$ occurs only when $w=1, x=0, y=0, z>Q_{\theta}$. Define $\delta_{17}$ as
\begin{align}
\delta_{17} &= \left\{
\begin{array}{ll}
1, & w=1, x=0, y=0, z>Q_{\theta}\\
&\textrm{and} \;\; (w, x, y, z) \in \mathbf{S}_{UTAB}\\
0, & \textrm{otherwise}.
\end{array}\right.
\label{UTAB_17}
\end{align}
The process moves from state $(w, x, y, z)$ to $(w+1, x, y,
z)$ with rate $\delta_{17}\mu_{s}$.

\item[\textbf{2)}] When the state of LAA small cell is OFF, if the queue is greater than $Q_{\theta}$, then the state of LAA small cell is switched to Sensing. Therefore, the transition from states $(w, x, y, z)$ to $(w+1, x, y, z)$ occurs only when $w=0, z>Q_{\theta}$. Define $\delta_{18}$ as
\begin{align}
\delta_{18} &= \left\{
\begin{array}{ll}
1, & w=0, z>Q_{\theta}\\
&\textrm{and} \;\; (w, x, y, z) \in \mathbf{S}_{UTAB}\\
0, & \textrm{otherwise}.
\end{array}\right.
\label{UTAB_18}
\end{align}
The process moves from state $(w, x, y, z)$ to $(w+1, x, y,
z)$ with rate $\delta_{18}\mu_{off}$.

\item[\textbf{3)}] When the state of the LAA small cell is Sensing, if a Wi-Fi packet or the queue occupies the queue is less than $Q_{\theta}$, then the state of the LAA small cell is switched to OFF. Therefore, the transition from states $(w+1, x, y, z)$ to $(w, x, y, z)$ occurs only when $w+1=1, x=0, y=1$ or $w+1=1, z=0$. Define $\delta_{19}$ as
\begin{align}
\delta_{19} &= \left\{
\begin{array}{ll}
1, & (w+1=1, x=0, y=1)\\
&\cup (w+1=1, z< Q_{\theta})\\
&\textrm{and} \;\; (w+1, x, y, z) \in \mathbf{S}_{UTAB}\\
0, & \textrm{otherwise}.
\end{array}\right.
\label{UTAB_19}
\end{align}
The process moves from state $(w+1, x, y, z)$ to $(w, x, y,
z)$ with rate $\delta_{19}\mu_{s}$.

The transitions between $(w, x, y, z)$ and $(w-1, x, y, z)$ is similar to that between $(w, x, y, z)$ and $(w+1,
x, y, z)$, and $\delta_{17}$, $\delta_{18}$,
\ldots , $\delta_{23}$ are obtained from~(\ref{UTAB_17}),~(\ref{UTAB_18}), \ldots ,~(\ref{UTAB_23}), respectively, and
\begin{align}
\delta_{21} &= \left\{
\begin{array}{ll}
1, & w-1=1, x=0, y=0, z> Q_{\theta}\\
&\textrm{and} \;\; (w-1, x, y, z) \in \mathbf{S}_{UTA}\\
0, & \textrm{otherwise}.
\end{array}\right.\\
\delta_{22} &= \left\{
\begin{array}{ll}
1, & w-1=0, z>Q_{\theta}\\
&\textrm{and} \;\; (w-1, x, y, z) \in \mathbf{S}_{UTA}\\
0, & \textrm{otherwise}.
\end{array}\right.\\
\delta_{23} &= \left\{
\begin{array}{ll}
1, & (w=1, x=0, y=1) \cup (w=1, z<Q_{\theta})\\
&\textrm{and} \;\; (w, x, y, z) \in \mathbf{S}_{UTA}\\
0, & \textrm{otherwise}.
\end{array}\right.
\label{UTAB_23}
\end{align}

Based on the description of the UFA in the previous section, when an
LAA packet arrives at the states where the buffer queue is full,
and there is no free unlicensed band server (i.e.,
$D_{f}=0, Q_{f}=0$), this LAA packet will be
dropped. On the other hand, there is no free unlicensed band server (i.e., $D_{f} = 0$), and the unlicensed band serving packet is LAA (i.e., $x \neq 0$); a Wi-Fi packet will be dropped. Therefore, the
dropping probabilities of the LAA and Wi-Fi for the UFA, UTA, UFAB, and UTAB are also expressed
as (\ref{block_prob_LAA}) and (\ref{block_prob_WF}).

From (\ref{Balance Equation}), (\ref{block_prob_LAA}) and
(\ref{block_prob_WF}), the steady state probabilities $\pi_{w,
x, y, z}$ and $P_{b, \ell}$ and $P_{b, w}$ can be computed by
the iterative algorithm in the section~\ref{sect:analytical_UFA} for LAA
coexisting with Wi-Fi heterogeneous networks.

\end{itemize}

\subsection{Simulation Validation}
\label{sect:validation}

\begin{table}[h]
\centering
\caption{Validation of the simulation and analysis
results for the LAA and Wi-Fi packets.}
\resizebox{1\columnwidth}{!}{
\begin{tabular}{|c|c|c|c|c|c|c|}
\hline
\multicolumn{7}{|c|}{$E[t_{w}] = \frac{1}{40}~\mbox{sec},\;
E[t_{\ell, u}] = \frac{1}{25}~\mbox{sec},\;
E[t_{on}] = E[t_{off}] = 10~\mbox{sec},\;
E[t_{s}] = 1~\mbox{sec},\; D=1,\; Q=2$}\\
\hline
\multicolumn{2}{|c|}{$\lambda_{\ell}$ (unit: $\mbox{user/sec}$)} & 25 & 37 & 50 & 62.5 & 120\\
\hline
\multicolumn{2}{|c|}{$\lambda_{w}$ (unit: $\mbox{user/sec}$)} & 5 & 5 & 5 & 5 & 5\\
\hline
$P_{b, \ell}$ of the UFA & Analytic   & 0.250425 & 0.409601 & 0.532753 & 0.614984 & 0.792439\\
                         & Simulation & 0.255031 & 0.412148 & 0.535449 & 0.616789 & 0.793422\\
                         & Error (\%) & 1.839273 & 0.621825 & 0.506051 & 0.293504 & 0.124047\\
\hline
$P_{b, w}$ of the UFA   & Analytic   & 0.745041 & 0.870437 & 0.931242 & 0.959145 & 0.992457\\
                        & Simulation & 0.743667 & 0.870636 & 0.929864 & 0.958482 & 0.991740\\
                        & Error (\%) & 0.001844 & 0.022862 & 0.147974 & 0.069124 & 0.072245\\
\hline
$P_{b, \ell}$ of the UTA & Analytic   & 0.452044 & 0.558498 & 0.646604 & 0.707167 & 0.840875\\
                         & Simulation & 0.415108 & 0.527295 & 0.624808 & 0.692863 & 0.841640\\
                         & Error (\%) & 8.170886 & 5.586949 & 3.370842 & 2.022719 & 0.090977\\
\hline
$P_{b, w}$ of the UTA   & Analytic   & 0.552184 & 0.657476 & 0.710252 & 0.734986 & 0.765312\\
                        & Simulation & 0.584350 & 0.698707 & 0.750545 & 0.766701 & 0.760551\\
                        & Error (\%) & 5.825232 & 6.271103 & 5.673057 & 4.315048 & 0.622099\\
\hline
\end{tabular}
}
\label{analysis_sim}
\end{table}

In this study, the analytic models are validated by simulation experiments. We adopt the FTP traffic model~\cite{3GPP36.814} as the packet traffic model destined to the UE.
Furthermore, allocations such as UFA, UTA, UFAB, and UTAB are evaluated by
simulation experiments without analytic modeling. The model follows
the discrete event-driven simulation approach in~\cite{Chou2019,Phone2001,Phone2003,YBLin2000}. We run $1,000,000$ application sessions in an experiment to ensure the convergence of the simulation results.
Table~\ref{analysis_sim} lists $P_{b, \ell}$ and $P_{b, w}$ values
for both analytic and simulation models for the UFA and UTA. The details of the
parameter setup in this table will be described in the following
section. Table~\ref{analysis_sim} indicates that the analytic results of the UFA match closely
with the simulation data. In Table~\ref{analysis_sim}, the errors between simulation and analytic
models of the UTA are below $8\%$ in most cases and are always less than
$6\%$. The fraction of time by UTA induces the same deviations between analytical and simulation under low load conditions.
Since the Markov analytical model has a memoryless property, and the unsteady state is caused by UTA, the analytical results cannot be accurately converged.
Especially when the ON/OFF state switching is more frequent, the error is more obvious.
The inherent lack of ON/OFF unsteady state by UTA, as well as imperfect convergence of the memoryless property, may induce some deviations between the analytical and simulation.
However, the overall trend of simulation and analysis results is still consistent.
Similar results for the UFAB and UTAB are observed and are not presented.
Furthermore, the conventional analytical models proposed in~\cite{Mehrnoush2018,Qixun2016,Gao2016} are based on full-buffering conditions and focus on throughput performance, so the performance results are ideal and there is no validation between simulation and analysis. This study is based on the finite-buffering condition of the limited dimension state analysis, which is more practical.

\section{Performance Evaluation}
\label{sect:performance}

This section investigates the performance of the unlicensed band allocations UFA, UTA, UFAB, and UTAB. In our
study, the input parameters $\lambda_{\ell}$, $\lambda_{w}$ and
$\mu_{\ell, u}$ are normalized by $\mu_{w}$. For example, if the
expected service time of a Wi-Fi packet is $1/\mu_{w} = 10$ seconds, then
$\lambda_{w} = \mu_{w}/2$ means that the expected Wi-Fi data packet
inter-arrival time at a cell is five seconds. To simplify our discussion, we assume that there is one unlicensed channel per cell, that is, $D = 1$.
For the cases where $D > 1$, similar results are observed and are not presented.
In this paper, the UFAB and UTAB are compared with the UFA and UTA to investigate how
the buffering mechanism for the packets in the UFAB and UTAB affects
the performance for both LAA and Wi-Fi packets. Our study indicates that for the LAA data, the acceptance rate and the effect of
Wi-Fi transmission, the UTAB outperforms other allocations
in most cases. This high acceptance rate is achieved
by reasonably slowing down the LAA packet transmission by the buffering mechanism.
And the ranges of buffer spaces $Q$ in most of the evaluations are assumed to be 1 to 5, and some evaluations are assumed to be 1 to 8 or 1 to 9 for more detailed observations.

\subsection{Performance of the UTA}
\begin{figure}[t]
\centering
\includegraphics[width=.43\textwidth]{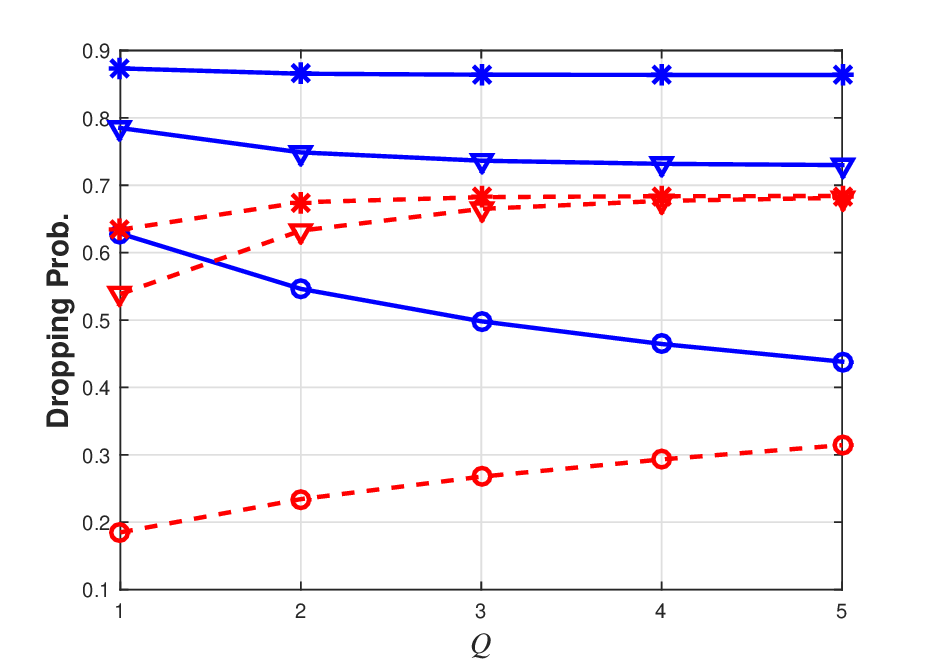}
\caption{Effects of the LAA packet arrival rates a on the LAA and Wi-Fi dropping probability in the UTA, where ``$\circ$'' represents $\lambda_{\ell} = \mu_{w}$, ``$\bigtriangledown$'' represents $\lambda_{\ell} = 5\mu_{w}$, and ``$\ast$'' represents $\lambda_{\ell} = 10\mu_{w}$.}
\label{LAA_arrival_rate}
\end{figure}

\begin{figure}[t]
\centering
\includegraphics[width=.43\textwidth]{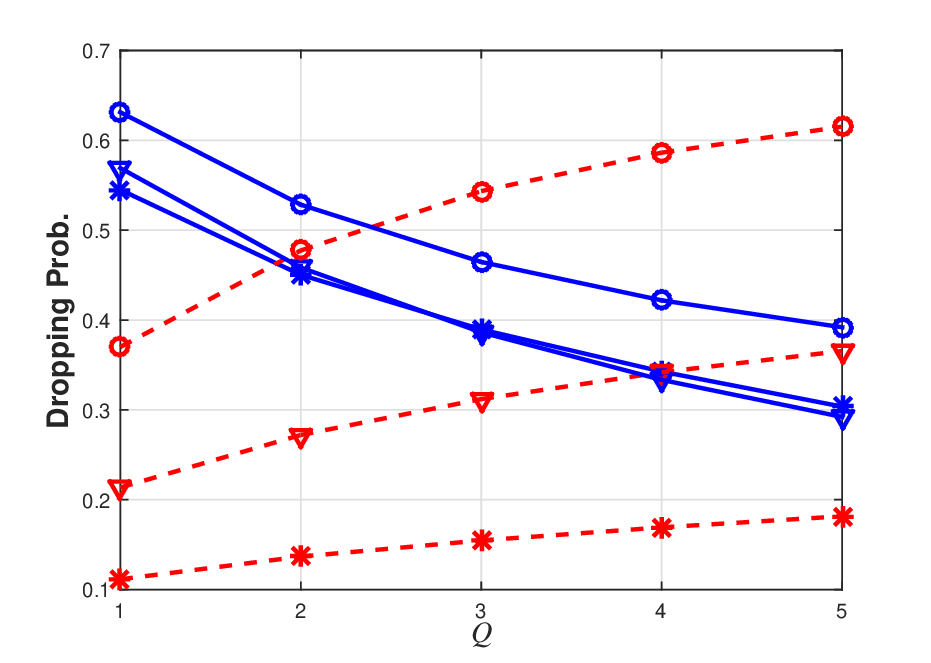}
\caption{Effects of the LAA packet service rates on the LAA and Wi-Fi dropping probability in the UTA, where ``$\circ$'' represents $\mu_{\ell,u} = 0.5\mu_{w}$, ``$\bigtriangledown$'' represents $\mu_{\ell,u} = \mu_{w}$, and ``$\ast$'' represents $\mu_{\ell,u} = 2\mu_{w}$.}
\label{LAA_service_rate}
\end{figure}

\begin{figure}[t]
\centering
\includegraphics[width=.43\textwidth]{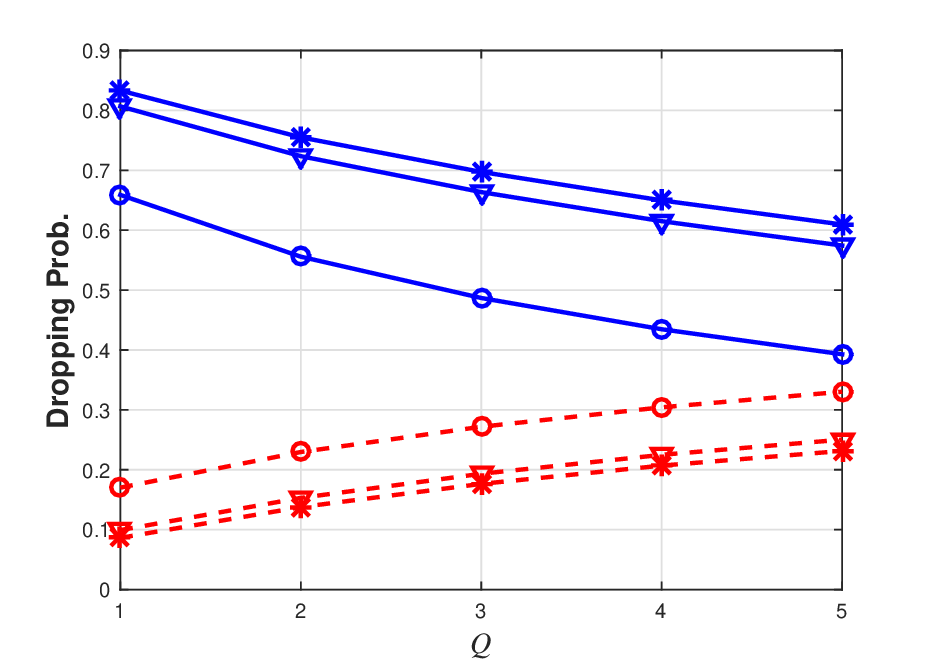}
\caption{Effects of the Wi-Fi packet arrival rates on the LAA and Wi-Fi dropping probability in UTA, where ``$\circ$'' represents $\lambda_{w} = \mu_{w}$, ``$\bigtriangledown$'' represents $\lambda_{w} = 5\mu_{w}$, and ``$\ast$'' represents $\lambda_{w} = 10\mu_{w}$.}
\label{WF_arrival_rate}
\end{figure}

\begin{figure}[t]
\centering
\includegraphics[width=.43\textwidth]{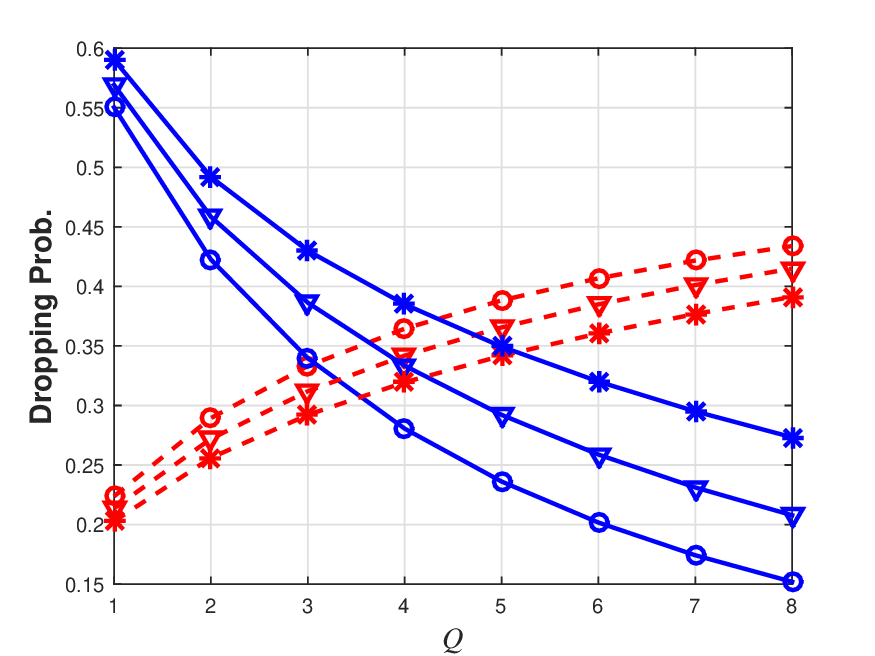}
\caption{Effects of the Wi-Fi packet service rates on the LAA and Wi-Fi dropping probability in the UTA, where ``$\circ$'' represents $\mu_{w} = 5$, ``$\bigtriangledown$'' represents $\mu_{w} = 10$, and ``$\ast$'' represents $\mu_{w} = 20$.}
\label{WF_service_rate}
\end{figure}

\begin{figure}[t]
\centering
\includegraphics[width=.43\textwidth]{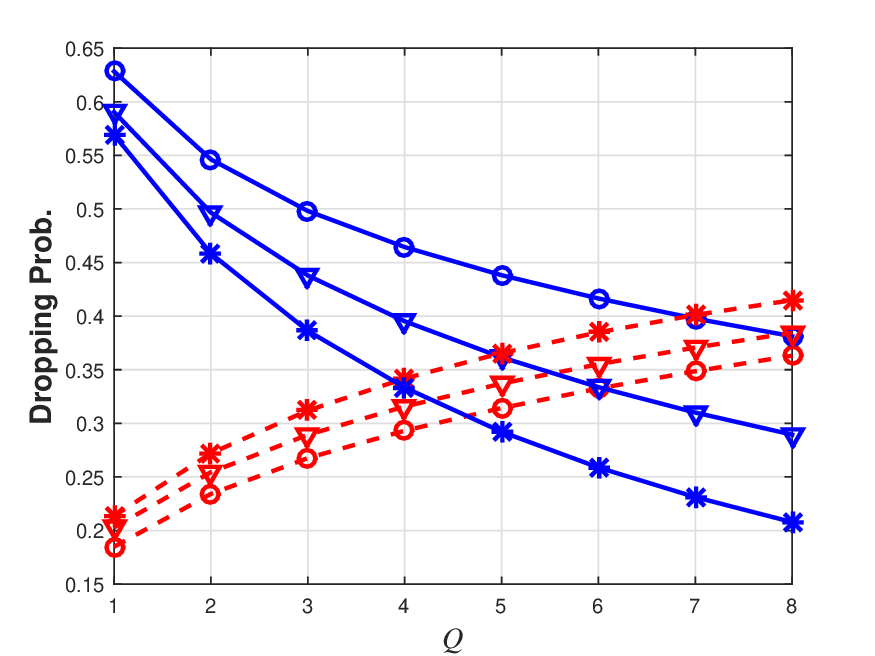}
\caption{Effects of the LAA sensing phase on the LAA and Wi-Fi dropping probability in UTA, where ``$\circ$'' represents $\mu_{s} = 0.1$, ``$\bigtriangledown$'' represents $\mu_{s} = 0.5$, and ``$\ast$'' represents $\mu_{s} = 1$.}
\label{sensing_phase}
\end{figure}

\begin{figure}[t]
\centering
\includegraphics[width=.43\textwidth]{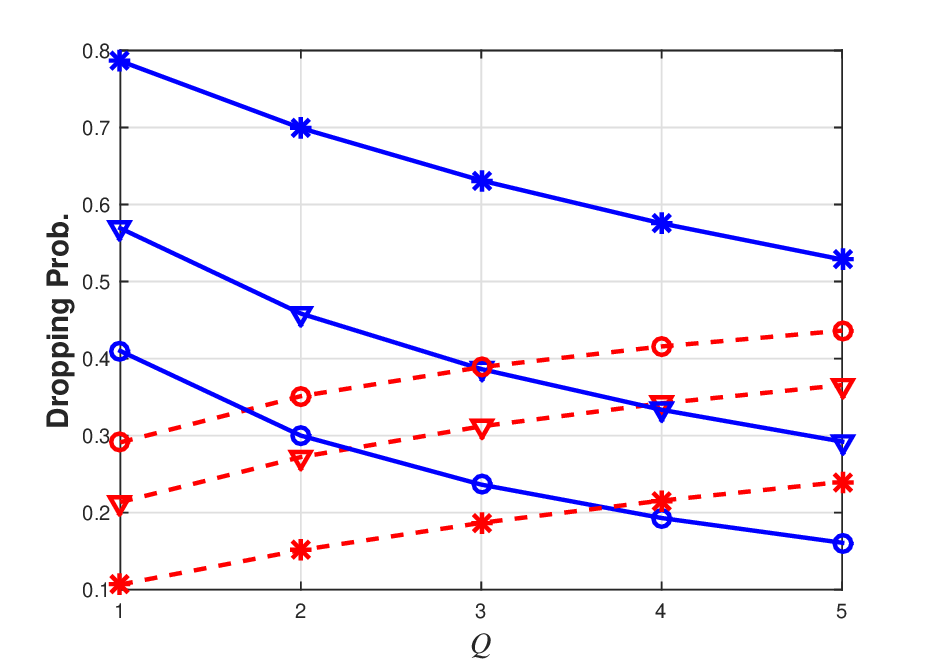}
\caption{Effects of the LAA channel occupation phase on the LAA and Wi-Fi dropping probability in the UTA, where ``$\circ$'' represents $\mu_{on} = 0.05\mu_{w}$ and $\mu_{off} = 0.2\mu_{w}$, ``$\bigtriangledown$'' represents $\mu_{on} = 0.1\mu_{w}$ and $\mu_{off} = 0.1\mu_{w}$, and ``$\ast$'' represents $\mu_{on} = 0.2\mu_{w}$ and $\mu_{off} = 0.05\mu_{w}$.}
\label{channel_occupation_phase}
\end{figure}

In this section, we use the UTA to investigate the performance of LAA
and Wi-Fi data transmission. The UTA has the LBT behavior (ON/OFF/Sensing states switch), and the UFA is a special
case of the UTA (Only ON state). The UFA and UTA have the same performance trends
in Fig.~\ref{LAA_arrival_rate}-\ref{channel_occupation_phase}, so we omit this part.
The effects of the input parameters are discussed
below, where the dashed (red) lines represent the Wi-Fi performances, and the solid (blue) lines represent the LAA performances.
Similar results were observed for other allocations and are omitted here for brevity.

\emph{1) Effects of LAA Packet Arrival Rate ($\lambda_{\ell}$):}

Fig.~\ref{LAA_arrival_rate} plots $P_{b,
\ell}$ and $P_{b, w}$ as a function of $\lambda_{\ell}$ and $Q$ for UTA, where $\lambda_{w} = 0.5\mu_{w}, \mu_{\ell, u} = \mu_{w}, \mu_{on} = \mu_{off} = 1, \mu_{s} = 10\mu_{on}$ and $1 \leq Q \leq 5$.
A general phenomenon in Fig.~\ref{LAA_arrival_rate} is that $P_{b, \ell}$ decreases and $P_{b, w}$ increases as $Q$ increases. This phenomenon
reflects the well-known result that the performance of a system (packet dropping probability) becomes better as the LAA packet arrival becomes bursty, since each packet requests more queue space. We also observe an intuitive result that $P_{b, \ell}$ and $P_{b, w}$ increases as $\lambda_{\ell}$ increases. However, $P_{b, w}$ will not increase without limit, because the UTA has the OFF phase and idle phase to guarantee the QoS of the Wi-Fi.

\emph{2) Effects of LAA Packet Size ($\mu_{u,\ell}$):}

Fig.~\ref{LAA_service_rate} plots $P_{b, \ell}$ and $P_{b, w}$ as a function of $\mu_{\ell}$ and $Q$ for the UTA, where $\lambda_{\ell} = 0.5\mu_{\ell}$ and the other input parameters are set the same as that in Fig.~\ref{LAA_arrival_rate}. Fig.~\ref{LAA_service_rate} shows the effects of the LAA packet sizes, where
$\mu_{\ell, u} = 2 \mu_{w}$ (small LAA packet size) and $\mu_{\ell, u}
= 0.5 \mu_{w}$ (large LAA packet size). A general phenomenon in Fig.~\ref{LAA_service_rate} is that $P_{b, w}$ increases as $Q$ increases. This result also
reflects the bursty LAA packet effect as observed in Fig.~\ref{LAA_arrival_rate}. We also observe an result that $P_{b, w}$ decreases significantly than $P_{b, \ell}$
as $\mu_{\ell , u}$ increases. This phenomenon implies that we should transmit the small-sized LAA packets on the unlicensed band.

\emph{3) Effects of Wi-Fi Packet Arrival Rate ($\lambda_{w}$):}

Fig.~\ref{WF_arrival_rate} plots $P_{b, \ell}$ and $P_{b, w}$ as a function of $\lambda_{w}$ and $Q$ for the UTA, where the other input parameters are set
the same as those in Fig.~\ref{LAA_arrival_rate}. Fig.~\ref{WF_arrival_rate} shows the effects of Wi-Fi packet arrival rates,
where $\lambda_{w} = \mu_{w}$ (small Wi-Fi packet arrival rate) and
$\lambda_{w} = 5\mu_{w}$ (large Wi-Fi packet arrival rate). The Wi-Fi traffic is large, and the LAA arrival packets become less competitive compared with the Wi-Fi packets. Thus, LAA packet requests have
less chance to be served as $\lambda_{w}$ increases, and $P_{b, w}$ decreases as $\lambda_{w}$ increases. The effect diminishes as $\lambda_{w}$ continues to increase.

\emph{4) Effects of Wi-Fi Packet Size ($\mu_{w}$):}

Fig.~\ref{WF_service_rate} plots $P_{b, \ell}$ and $P_{b, w}$ as a function of $\mu_{w}$ and $Q$ for the UTA, where $1 \leq Q \leq 8$ and the other input parameters are set
the same as that in Fig.~\ref{LAA_service_rate}. Since $\mu_{w}$ increases, the small Wi-Fi packets have a better chance of being served on the unlicensed band. Thus, $P_{b, w}$ decreases and $P_{b, \ell}$ increases as $\mu_{w}$ increases. Otherwise, $P_{b, \ell}$ decreases significantly as $\mu_{w}$ increases, which confirms the feasibility of TV white space application. However, if the free spaces of the queue ($Q$) are enough, it is easier to achieve the balance performance between the LAA and Wi-Fi that means that $P_{b,l}$ and $P_{b,w}$ are equivalent blocking probabilities (see $Q = 3, 4$, and $5$ in Fig.~\ref{WF_service_rate}). Therefore, with the larger Wi-Fi packet size, it is necessary to increase the number of queuing spaces appropriately, but be aware that extremely more or fewer queuing spaces will make the performance unbalanced between the LAA and Wi-Fi.

\emph{5) Effects of LBT ($\mu_{s}$, $\mu_{on}$ and $\mu_{off}$):}

Fig.~\ref{sensing_phase} plots $P_{b, \ell}$ and $P_{b, w}$ as a function of $\mu_{s}$ and $Q$ for the UTA, where $\lambda_{\ell} = \lambda_{w} = 0.5\mu_{w}, \mu_{\ell, u} = \mu_{w}, \mu_{on} = \mu_{off} = 0.1\mu_{s}, D = 1$ and $1 \leq Q \leq 8$. In the real world, the sensing accuracy of the LBT is increased as $\mu_{s}$ decreases. However, the $\mu_{s}$ decreased, and the unlicensed spectrum efficiency of the LAA system will be degraded. Therefore, we observe that
$P_{b, w}$ decreased and $P_{b, \ell}$ significantly increased as $\mu_{s}$ decreased. Similar to Fig.~\ref{WF_service_rate}, when the idle phase is longer, we need to increase the number of queue spaces for buffering appropriately. Adjusting the number of free spaces in the queue will assist $P_{b, \ell}$ and $P_{b, w}$ to achieve the balanced performance between the LAA and Wi-Fi.

Fig.~\ref{channel_occupation_phase} plots $P_{b, \ell}$ and $P_{b, w}$ as a function of $\mu_{on}$, $\mu_{off}$ and $Q$ for the UTA, where $\lambda_{\ell} = \lambda_{w} = 0.5\mu_{w}, \mu_{\ell, u} = \mu_{w}, \mu_{s} = \mu_{w}, D = 1$ and $1 \leq Q \leq 5$. A general phenomenon in Fig.~\ref{channel_occupation_phase} is that $P_{b, \ell}$ decreased and $P_{b, w}$ increased as $\mu_{on}$ increased and $\mu_{off}$ decreased, vice versa.
But adjusting the number of free spaces in the queue will assist $P_{b, \ell}$ and $P_{b, w}$ to achieve the balanced performance between the LAA and Wi-Fi.
In~\cite{3GPP2015}, the lengths of ON duration ($t_{on}$) and OFF duration ($t_{off}$) must be the same length ($t_{on} = t_{off}$) and require sufficient buffer space to achieve the performance balance between the LAA and Wi-Fi (See $Q = 4$ in Fig. 10). Under the same conditions, when the lengths of ON duration and OFF duration are allowed to be different, if $t_{on} >t_{off}$, the buffer space only needs $Q = 2$ to achieve the performance balance between LAA and Wi-Fi, it can save two more buffer spaces than the case of $t_{on} = t_{off}$ for Fig.~\ref{channel_occupation_phase} in this paper. Conversely, if $t_{on} <t_{off}$, even though the buffer space $Q$ is increased to 5, the performance balance between LAA and Wi-Fi cannot be achieved.

\subsection{Comparison for the UFA, UTA, UFAB, and UTAB }

\begin{figure}[t]
\centering
\includegraphics[width=.43\textwidth]{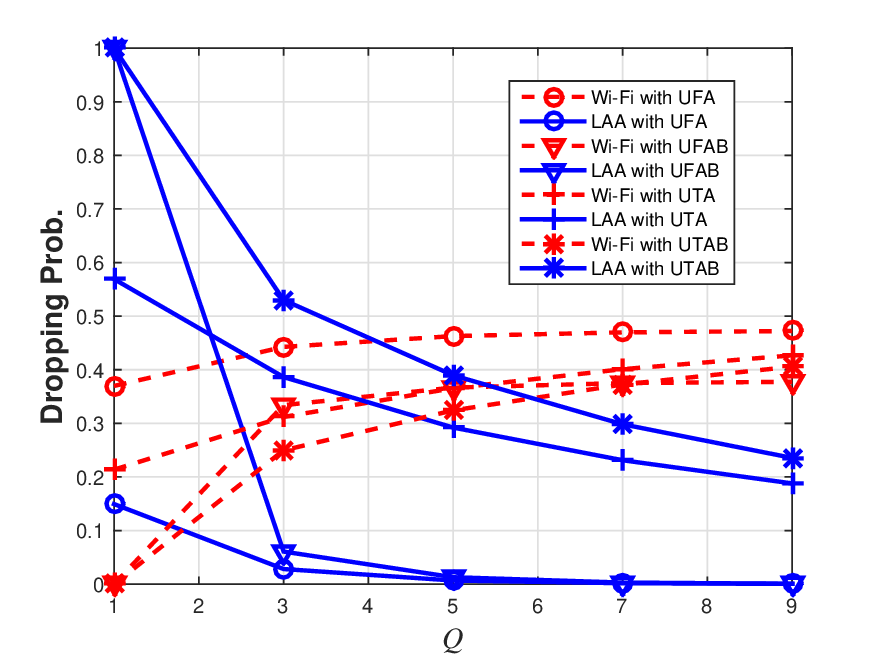}
\caption{Impact of the LAA queue size on the LAA and Wi-Fi dropping probability.}
\label{queue_size}
\end{figure}

This section compares the performance for the UFA, UTA, UFAB, and UTAB.

\emph{1) Effects of the Queue Size ($Q$):}

Fig.~\ref{queue_size} plots $P_{b, \ell}$ and $P_{b, w}$ as a function of $Q$ for the UFA, UTA, UFAB and UTAB.
In the Fig.~\ref{queue_size}, we set $\lambda_{\ell} = \lambda_{w} = 0.5\mu_{w}, \mu_{\ell, u} = \mu_{w}, \mu_{on} = \mu_{off} = 0.1\mu_{w}, \mu_{s} = \mu_{w}, D = 1$ and $1 \leq Q \leq 9$.
For various input parameter setups, we observe the same results that will not be presented in this paper.
Fig.~\ref{queue_size} shows that the UFA outperforms the other allocations in terms of the $P_{b, \ell}$ performance, and the UTAB outperforms the other allocations in terms of the $P_{b, w}$ performance.
We observe that the UFAB and UTAB outperform the UFA and UTA in terms of the packet dropping probability $P_{b, w}$, which implies that the buffering mechanism for LAA packets significantly reduces the Wi-Fi packet dropping probability. In the UFA, the performance between the LAA and Wi-Fi has significantly differed and cannot be compensated for by any mechanism, which is the worst performance among all schemes. In the UFAB, the performance of the LAA and Wi-Fi can be improved by the buffering mechanism, but there is still a large performance difference between the LAA and Wi-Fi. The UTA has effectively improved the weak point of the UFAB. In the UTA, there is a significant balance between the LAA and Wi-Fi performance, especially for $Q = 3, 4$, and $5$. However, after $Q = 6$, the performance between the LAA and Wi-Fi still differs significantly. In UTAB, the performance between LAA and Wi-Fi is much more balanced than the UTA, including $Q = 5, 6$, and $7$, and the range of performance balance can be determined by the adjustment of the threshold $Q_{\theta}$. In summary, when the buffer size is small ($2 \leq Q \leq 4$), the improvements of $P_{b, \ell}$ and $P_{b, w}$ for the UTA are the best choice, and when the buffer size is large ($5 \leq Q \leq 7$), to maintain both QoS for the LAA and Wi-Fi packet data users, the UTAB is the better choice than the UTA. Therefore, the buffering mechanism is the necessary setting for LAA systems.

\emph{2) Effects of the Buffering Mechanism ($Q_{\theta}$):}
\begin{figure}[t]
\centering
\includegraphics[width=.43\textwidth]{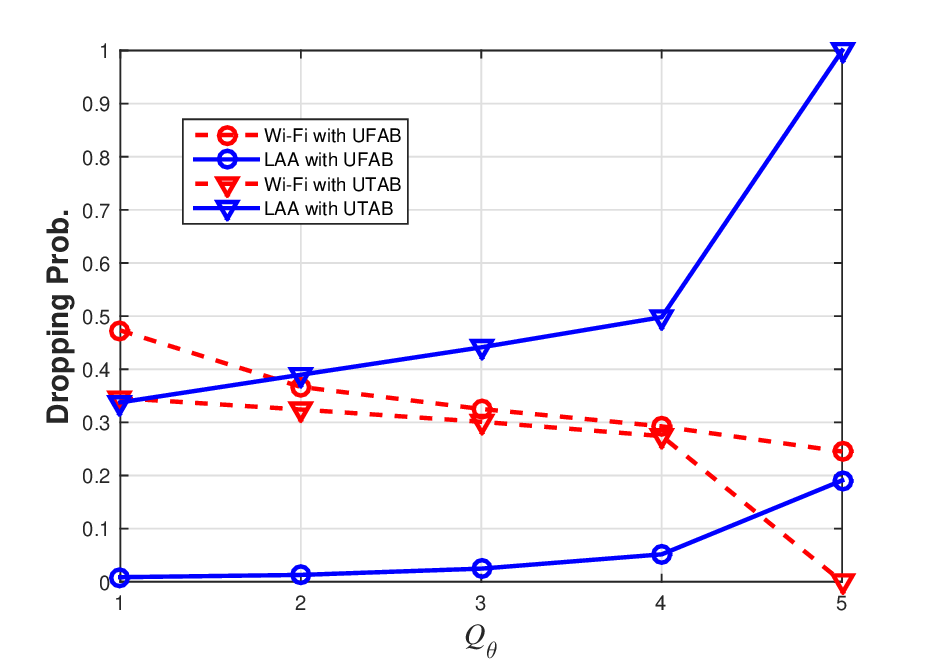}
\caption{Impact of the LAA buffer threshold on the LAA and Wi-Fi dropping probability.}
\label{buffer_threshold}
\end{figure}

Fig.~\ref{buffer_threshold} plots $P_{b, \ell}$ and $P_{b, w}$ as a function of $Q_{\theta}$ for the UFAB and UTAB.
In the Fig.~\ref{buffer_threshold}, we set $\lambda_{\ell} = \lambda_{w} = 0.5\mu_{w}, \mu_{\ell, u} = \mu_{w}, \mu_{on} = \mu_{off} = 0.1\mu_{w}, \mu_{s} = \mu_{w}, D = 1, Q = 5$ and $1 \leq Q_{\theta} \leq 5$.
For $Q_{\theta} \leq 4$, $P_{b, w}$ of the UFAB and UTAB are not affected by the buffer threshold $Q_{\theta}$.
When $Q_{\theta} = 1 $ and $2$, the UTAB have better balance between $P_{b, w}$ and $P_{b, \ell}$.
When $Q_{\theta} = 5$, the UFAB have better balance between $P_{b, w}$ and $P_{b, \ell}$.

\section{Conclusion}
\label{sect:conclusion}
This paper studied the impact of LAA transmission on the Wi-Fi network with the assumption of resource sharing.
Specifically, we proposed analytic and simulation models to investigate the performance of LAA and Wi-Fi networks.
The proposed model considers the interaction between LAA and Wi-Fi by using a global system state and a waiting queue that buffers the LAA packets when no channel is available.
We considered two LAA unlicensed band allocations (the UFA and UTA) as specified in the LAA standard and the corresponding buffering mechanism (the UFAB and UTAB).
Simulations were used to corroborate the analytical results, as close agreement between the two is observed.
Our study indicated that the UFA allocation effectively increases the LAA packet acceptance rate, the UTA allocation significantly reduces the Wi-Fi dropping probability, and the dropping probability of LAA and Wi-Fi packets can be adjusted by the buffer mechanism. The buffering mechanism significantly reduces both the LAA and Wi-Fi packet dropping probability.
In general, UFA outperforms the other allocations in terms of the $P_{b, \ell}$, and the UTAB outperforms the other allocations in terms of the $P_{b, w}$.
From the perspective of balancing LAA and Wi-Fi, UFAB outperforms the other allocations when the buffer threshold is large ($Q_{\theta} = 5$, where $Q=5$), and the UTAB outperforms the other allocations in the case of the small buffer threshold ($Q_{\theta} = 2$, where $Q=5$). Finally, our study can provide guidelines for the operators to set the ON/OFF timer duration and the buffering size of the LAA small cell to reduce the packet dropping and the impact of buffer overflow by properly selecting the buffer threshold.

\bibliographystyle{IEEEtran}

\vspace{-0.2in}

\begin{IEEEbiography}[{\includegraphics[width=1in,height=1.25in,clip,keepaspectratio]{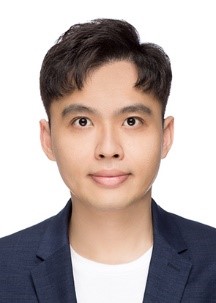}}]{Po-Heng Chou} (Member, IEEE) was born in Tainan, Taiwan. He received the B.S. degree in electronic engineering from National Formosa University (NFU), Huwei, Yunlin, Taiwan, in 2009, the M.S. degree in communications engineering from National Sun Yatsen University (NSYSU), Kaohsiung, Taiwan, in 2011, and the Ph.D. degree from the Graduate Institute of Communication Engineering (GICE), National Taiwan University (NTU), Taipei, Taiwan, in 2020. His research interests include AI for communications, deep learning-based signal processing, wireless networks, and wireless communications, etc.

He was a Postdoctoral Fellow at the Research Center for Information Technology Innovation (CITI), Academia Sinica, Taipei, Taiwan, from Sept. 2020 to Sept. 2024. 
He was a Postdoctoral Fellow at the Department of Electronics and Electrical Engineering, National Yang Ming Chiao Tung University (NYCU), Hsinchu, Taiwan, from Oct. to Dec. 2024.
He has been elected as the Distinguished Postdoctoral Scholar of CITI by Academia Sinica from Jan. 2022 to Dec. 2023. He is invited to visit Virginia Tech (VT) Research Center (D.C. area), Arlington, VA, USA, as a Visiting Fellow, from Aug. 2023 to Feb. 2024.
He received the Partnership Program for the Connection to the Top Labs in the World (Dragon Gate Program) from the National Science and Technology Council (NSTC) of Taiwan to perform advanced research at VT Institute for Advanced Computing (D.C. area), Alexandria, VA, USA, from Jan. 2025 to present.

Additionally, Dr. Chou received the Outstanding University Youth Award and the Phi Tau Phi Honorary Membership from NTU in 2019 to honor his impressive academic achievement. He received the Ph.D. Scholarships from the Chung Hwa Rotary Educational Foundation from 2019 to 2020.
\end{IEEEbiography}

\end{document}